\documentclass[aps,preprint]{revtex4}
\usepackage{amsmath,amssymb,amsthm,mathrsfs,amsfonts,dsfont} 
\usepackage{graphicx}
\usepackage{subfigure}
\usepackage{amsbsy}
\usepackage{bm}
\usepackage{xcolor}

\usepackage{algorithm}
\usepackage{algpseudocode}
\usepackage{algorithmicx}

\newcommand{\blue}[1]{{\color{black}#1}}

\def\bx{\mathbf{x}}

\begin{document}

\title{Synaptic plasticity alters the nature of chaos transition in neural networks}
\author{Wenkang Du$^{1}$}
\author{Haiping Huang$^{1,2}$}
\email{huanghp7@mail.sysu.edu.cn}
\affiliation{$^{1}$PMI Lab, School of Physics,
Sun Yat-sen University, Guangzhou 510275, People's Republic of China}
\affiliation{$^{2}$Guangdong Provincial Key Laboratory of Magnetoelectric Physics and Devices,
Sun Yat-sen University, Guangzhou 510275, People's Republic of China}
\date{\today}

\begin{abstract}
	In realistic neural circuits, both neurons and synapses are coupled in dynamics with separate time scales. The circuit functions are intimately related to these coupled dynamics. However, it remains challenging to understand the intrinsic properties of the coupled dynamics. Here, we develop the neuron-synapse coupled quasi-potential method to demonstrate how learning induces a qualitative change in the macroscopic behaviors of recurrent neural networks. We find that under the Hebbian learning, a large Hebbian strength will alter the nature of the chaos transition, from a continuous type to a discontinuous type, where the onset of chaos requires a smaller synaptic gain compared to the non-plastic counterpart network. In addition, our theory predicts that under feedback and homeostatic learning, the location and type of chaos transition are retained, and only the chaotic fluctuation is adjusted. Our theoretical calculations are supported by numerical simulations. 
\end{abstract}

 \maketitle

\section{Introduction}
Revealing mechanisms underlying brain dynamics is one of the most fascinating scientific endeavors of this century. Brain dynamics support our thinking, perception and memory~\cite{RMP-2006,ND-2014}, involving two types of coupled dynamic processes---neuronal and synaptic dynamics. The synaptic dynamics control how neurons are non-reciprocally connected, yielding complex neuronal dynamics (e.g., self-organized criticality in neocortex~\cite{SOC-2007}), while the synaptic connections are in turn affected by the evolving neural states (namely synaptic plasticity)~\cite{Plast-2020,Abbott-2000}. These two sides of brain dynamics complicated all theoretical analyses in previous works~\cite{Amari-1977,WCM-1972,Huang-2024}, making a complete understanding of the role of synaptic plasticity still challenging, \blue{despite some recent progress on associative memory models with supervised or unsupervised Hebbian learning~\cite{NN-2024,EPL-2023}.}

Physics has a long history of studying spin dynamics. At a coarse-grained level, the spin state can be treated as a neuronal state. 
The path integral approach or dynamical mean-field theory was first introduced to study spin dynamics trajectories~\cite{Domin-1978,Sommers-1987,Haim-1987} in models where the spin
couplings are randomly quenched. However, a later development involved relatively slow dynamics of spin couplings~\cite{JPA-1993,JPA-1994}, marking an important step both in concepts and techniques towards understanding a complex system of coupled dynamics. On the neural dynamics side, there emerged a lot of 
theoretical studies about dynamical behaviors of randomly connected neural networks, such as chaos transitions found in high dimensional neural dynamics~\cite{Chaos-1988,Qiu-2024,PRL-2013,Helias-2022,Wang-2024}, random synapses of low-rank structures~\cite{Ostojic-2020},  and stochastic nonlinear
neuronal dynamics with background noises~\cite{Helias-2020b}. Recent works started to combine both dynamics in a machine learning system~\cite{Hinton-2016,Jstat-2023},
and it was recently addressed how the random untrained and trained parts of the couplings interact to produce expected performances~\cite{Barak-2020}, and it was also demonstrated how a Hebbian hierarchy affects retrieval dynamics of memory sequences~\cite{Jiang-2023}. In particular, the dynamical mean-field theory was recently used to study the coupled system and reveal that synaptic dynamics can speed up or slow down neuronal dynamics, and thus the chaos can be made freezable (akin to a working memory function)~\cite{Clark-2024}. An intrinsic time scale introduced to Hebbian coupling dynamics leads to the result that older memory and recent memory bear different chaotic temporal fluctuations~\cite{Brunel-2023}. Therefore, studying the coupled dynamics based on the dynamical mean-field theory becomes an active scientific frontier in theoretical neuroscience. 

The dynamical mean-field analysis would become very complicated in reduced descriptions of the high-dimensional coupled dynamics, as partial differential equations need to be solved, which prevents us from studying those complicated scenarios of synaptic and neuronal dynamics and further plasticity-induced phase transitions. Inspired by recent works of quasi-potentials for non-equilibrium neural dynamics~\cite{Qiu-2024}, we propose a canonical ensemble theory to address the interplay between synaptic plasticity and neural dynamics, focusing only on their zero-speed (or fixed point) limit. We consider three types of commonly used plasticity rules---Hebbian plasticity, random feedback-driven learning, and target rate-oriented homeostatic plasticity, which all bear machine learning benefits~\cite{Abbott-2012,David-1994} and neurobiological relevance~\cite{Hebbian-1999,Homeo-1999,Homeo-2022,Rate-2016}. Surprisingly, we reveal that the plasticity parameter tunes the nature of the chaos transition from the first order to the second order, which will significantly impact the intrinsic structure of the phase space. We will detail the adopted methodology and discuss the scientific contribution to our understanding of the collective dynamical behavior of the coupled systems. 

\section{Equilibrium theory of learning}
In this section, we first introduce the recently proposed quasi-potential method for non-gradient neural dynamics, and then describe 
in detail the proposed framework to treat the theory of learning in this paper, \blue{in a different form from traditional replica theory of learning in associative memory models~\cite{NN-2024,EPL-2023}}. Our framework concentrates on the fixed-point limit of the dynamics, thereby avoiding solving dynamical mean field equations in the traditional path integral framework~\cite{Zou-2024,Roudi-2017,Chow-2015}. We shall show the advantage of the quasi-potential method in capturing dynamical phase transitions in this section.

\subsection{Quasi-potential method for non-equilibrium dynamics}
We consider a canonical model of an $N$-neuron coupled recurrent neural network (RNN), where the state of the network is described by the synaptic current $x_i(t)\in\mathbb{R}$. This synaptic current satisfies the following $N$-dimensional ordinary differential equation:
\begin{equation}\label{dyn}
  \frac{\mathrm{d} x_i}{\mathrm{~d} t}=-x_i+\sum_{j=1}^N J_{i j} \phi\left(x_j\right),
\end{equation}
where we exclude the self-coupling ($J_{ii}=0$), the first term on the right-hand side of the equation represents a natural decay in the absence of feedback inputs ($\mathbf{J}=\mathbf{0}$), while the second term denotes the influence of other neurons $j$ on neuron $i$, with the non-reciprocal coupling $J_{i j}\neq J_{j i}$. The function $\phi$ is a nonlinear activation function, which, in this paper, is assumed to be the $\tanh$ function by default. Each coupling is generated independently as $J_{ij}\sim\mathcal{N}(0,g^2/N)$, where $g$ captures the strength of synaptic feedback. 

The collective dynamical behavior was theoretically clarified in the seminal work~\cite{Chaos-1988}. The critical value of $g_c=1$ separates a trivial null-activity phase from a non-trivial chaotic phase, where two initially close trajectories will finally deviate with a positive rate (called Lyapunov exponent). The chaos transition was recently revealed to have a connection to the concept of topological complexity~\cite{PRL-2013,Helias-2022} and peaked response functions at the edge of chaos~\cite{Qiu-2024}, where the continuous nature of the chaos transition is mathematically justified. We next briefly introduce the quasi-potential method that we shall extend to address the neuronal-synaptic dynamics.

Focusing on the steady fixed-point (may be unstable) state of the non-gradient recurrent dynamics [Eq.~\eqref{dyn}], we can intuitively write down the following cost for optimization:
\begin{equation}
  E(\bx)=\frac{1}{2} \sum_i\left(-x_i+\sum_j J_{i j} \phi\left(x_j\right)\right)^2+\eta\|\bx\|^2,
\end{equation}
where the first term of the energy function represents the kinetic energy (considering the unit mass), while the second term is the regularization term and $\eta$ is a predefined parameter. This optimization of continuous variables can be done by gradient dynamics with a certain level of white noise whose variance is determined by a temperature $T$.  It is then well known that the steady state of the stochastic gradient dynamics can be described by the Boltzmann distribution~\cite{Zou-2024}:
\begin{equation}
  P(\mathbf{x})=\frac{1}{Z} e^{-\beta E(\mathbf{x})},
\end{equation}
where $Z$ is the partition function, $\mathbf{x}$ represents the activity vector, and $\beta=1 / T$. By sending $\beta\to\infty$, we will immediately arrive at the zero speed limit,
i.e., all fixed points (regardless of their stability) can be captured under this Boltzmann measure. This constructs the core idea of the quasi-potential function, i.e., order parameters describing the non-equilibrium dynamics in the zero-speed limit can be obtained from the disorder average over $\mathbf{J}$, demonstrating that the order can emerge from apparent disorders. Qualitative behavior of the collective dynamics in the long-time limit can then be determined by these order parameters. Next, we will develop a theoretical framework to incorporate the effect of learning on neural dynamics. In other words, our goal is to address a fundamental question of how learning induces the change of the phase space structure underlying the high-dimensional chaotic dynamics~\cite{Huang-2024}.

\subsection{Canonical ensemble theory of learning}
We consider three types of synaptic plasticity rules. The first one is the standard Hebbian type, i.e., the pre-synaptic and post-synaptic neural activities affect 
the synaptic strength in the following way:
\begin{equation}\label{heb}
J_{i j}=J_{i j}^0+\frac{k}{N} \phi\left(x_i\right) \phi\left(x_j\right),
\end{equation}
where the first term indicates the untrained random substrate (e.g., generating the spontaneously chaotic fluctuations), while the second term 
explains the local Hebbian effect~\cite{Hebbian-1999}. The random substrate is a random matrix whose entries are independently generated from $\mathcal{N}\left(0, \frac{g^2}{N}\right)$, and $k$ specifies the strength of the Hebbian term. By varying the value of $k$, one can see how strongly the Hebbian term
affects the steady fixed-point state of the original random RNN.

The second plasticity we consider is the feedback learning~\cite{Abbott-2009}. More precisely, in reservoir computing, only a linear readout 
is trained, and then the output is sent via feedback to all units in the neural reservoir where the coupling is random and untrained (in analogy to
$\mathbf{J}^0$ here). The feedback weight is commonly chosen to be random as well, e.g., following a standard Gaussian distribution. In principle,
adding
 a feedback loop is a highly flexible way of increasing functional adaptability through learning, since nervous systems 
 often seem to be composed of loops~\cite{Abbott-2009}. For example, top-down attention may be sent back through this feedback loop. Interestingly,
 despite unchanged $\mathbf{J}^0$, the random feedback amounts to the equivalent rank-one modification of the untrained $\mathbf{J}^0$~\cite{Abbott-2012}, when the readout weight is updated to match the target signal. Therefore, we propose a toy model of this sort of feedback learning.
 \begin{equation}\label{fb}
J_{i j}=J_{i j}^0+\frac{\delta}{N} u_i \phi\left(x_j\right),
\end{equation}
where the constant $\delta$ characterizes the feedback strength, and the feedback weight $u_i$ is independently sampled from $\mathcal{N}(0,1)$. In essence, $\delta$ is related to the readout error. For example, optimizing a mean-squared error between actual and target outputs, the readout weight dynamics will approach the steady state $w_i^*=-\frac{\epsilon_\mu}{\lambda}\phi(x_i^\mu)$, where $\mu$ is the current training example, $\epsilon_\mu$ is the readout error, and $\lambda$ is the weight-decay parameter. Hence, the last term in Eq.~\eqref{fb} captures the rank-one perturbation $\Delta\mathbf{J}=\mathbf{u}\mathbf{w}^\top$~\cite{Abbott-2012}.

The third plasticity is the well-known homeostatic plasticity. The homeostatic plasticity is an important partner of Hebbian one, discovered in
cortical networks~\cite{Homeo-1999}. It commonly includes two types---multiplicative scaling of synaptic strength and activity-dependent stabilization of synaptic connections. Both types allow the network to satisfy the joint requirement of adaptation and stability. The latter one can be controlled by sleep and wake brain states~\cite{Rate-2016} and was further explored to support stable self-sustained dynamics~\cite{Homeo-2022}. Hence, in this work, we consider the activity-dependent one, namely, neuronal firing rate homeostasis:
\begin{equation}\label{hs}
    J_{i j}=J_{i j}^0-\frac{k}{N}\left[ \phi\left(x_i\right)-r_{\operatorname{tg}}\right] \phi\left(x_j\right),
\end{equation}
where $r_{\text {tg }}$ is the homeostatic setpoint of firing rates, and $k$ indicates a learning rate. In all three plasticity rules, the rank-one modification at the single synapse level is negligible but produces a meaningful impact on the global behavior of the network~\cite{Clark-2024} (see also our following replica calculations). Note also that the modification is asymmetric in the latter two forms of plasticity, in contrast to the Hebbian one, which could slow or suppress the chaotic fluctuation.

Taking into account the aforementioned plasticity rules, we write the fixed-point distribution under learning as an optimization:
\begin{equation}
  E_{\ell}(\bx)=\frac{1}{2} \sum_i\left(-x_i+\sum_j [J_{i j}^0+\Delta J_{ij}] \phi\left(x_j\right)\right)^2+\eta\|\bx\|^2,
\end{equation}
where $\Delta J_{ij}$ is the aforementioned activity-dependent modification to the random untrained part $J_{ij}^0$. Therefore, the fixed-point state follows a Boltzmann
distribution $P(\bx)=\frac{1}{Z_\ell}e^{-\beta E_\ell(\bx)}$, where the learning related partition function $Z_{\ell}$ is a central quantity in this paper.

We finally remark that $\Delta J_{ij}$ is the stationary solution of the following synaptic dynamics:
\begin{equation}
(1+\tau\partial_t)\mathbf{L}(t)=\Delta\mathbf{J}(t),
\end{equation}
where the synaptic time scale $\tau>1$ (in a biological plausible sense that the plasticity dynamics is slower than the neural dynamics),  the time-dependent plasticity $\mathbf{L}(t)$ is added to the random substrate $\mathbf{J}^0$ to form an intact $\mathbf{J}$, which further impacts the neural dynamics [Eq.~\eqref{dyn}].  In simulations, we shall support our theoretical results with the numerical solutions of the above coupled dynamics.

\subsection{Emergence of order from apparent disorder}
Now, we clarify the steps to observe an emergence of order from apparent disorder in the network coupling. For the canonical ensemble of learning, we must
compute the free energy function $f\equiv-T\ln Z_{\ell}$, where we take the unit Boltzmann constant as commonly adopted in statistical mechanics of optimization problems~\cite{MM-2009}, and clearly the partition function depends on the coupling realization. Hence, to obtain universal properties of the free energy, one
has to calculate the disorder-averaged free energy, which is tractable due to the following replica trick~\cite{Mezard-1987,Huang-2022}:
\begin{equation}
  -\beta f=\frac{1}{N}\langle\ln Z_\ell(\mathbf{J})\rangle_{\mathbf{J}} =\lim _{n \rightarrow 0} \frac{1}{n N} \ln \langle Z_\ell^n(\mathbf{J})\rangle_{\mathbf{J}},
\end{equation}
which corresponds to taking the vanishing rate function in the large deviation principle~\cite{PRL-2008}. The average over $\mathbf{J}$ may also include the randomness of the feedback weights for the feedback learning. The replicated partition function reads,
\begin{equation}
  \begin{aligned}
  Z^n_\ell(\mathbf{J}) & =\int d \mathbf{x} \exp \left[-\beta\left(\frac{1}{2} \sum_{i,a}\left(-x_i^a+\sum_j J_{i j} \phi\left(x_j^a\right)\right)^2+\eta \sum_a\left\|\mathbf{x}^a\right\|^2\right)\right], \\
  & =\int d \mathbf{x} D \hat{\mathbf{x}} \exp \left[i \sqrt{\beta} \sum_{i,a} \hat{x}_i^a\left(-x_i^a+\sum_j J_{i j} \phi\left(x_j^a\right)\right)-\beta \eta \sum_a\left\|\mathbf{x}^a\right\|^2\right],
  \end{aligned}
\end{equation}
where $J_{ij}=J_{i j}^0+\Delta J_{ij}$, $d\bx\equiv\prod_{i=1}^N\prod_{a=1}^ndx_i^a$, $D \hat{\mathbf{x}} \equiv \prod_{i=1}^N \prod_{a=1}^n D \hat{x}_i^a$ where $D \hat{x} \equiv e^{-\frac{1}{2} \hat{x}^2} d \hat{x} / \sqrt{2 \pi}$ is the Gaussian measure introduced during the linearization of the quadratic term through the Hubbard-Stratonovich transformation. We also introduce the replica index $a$, which runs from 1 to $n$, and the subscript $i$ represents the neuron index, which runs from $1$ to $N$.

After performing the quenched disorder average $\langle\cdot\rangle_{\mathbf{J}}$ (details provided in Appendix~\ref{app-a}), we naturally introduce two physically meaningful order parameters: 
\begin{equation}
\begin{aligned}
Q^{a b} & =\frac{1}{N} \sum_i \phi\left(x_i^a\right) \phi\left(x_i^b\right), \\
R^a & =\frac{1}{N} \sum_i \hat{x}_i^a \phi\left(x_i^a\right).
\end{aligned}
\end{equation}
The first order parameter characterizes the fluctuation of the neural firing rates, a plausible quantity for detecting phase transitions, while the second order parameter corresponds to the response function in statistical physics, since it is in fact the derivative of the mean population activity with respect to external perturbation of a small current~\cite{Qiu-2024}. In this sense, the abstract overlap matrix ($\mathbf{Q}$ and $\mathbf{R}$) can be linked to the measurable neural activity in the complex dynamical system. At the first level of approximation, which can be cross-checked by experiments and the stability of the resultant mean-field equations, we write the following replica symmetric (RS) ans\"atz,
\begin{equation}
\begin{aligned}
Q^{a b} & =q \delta_{a b}+Q\left(1-\delta_{a b}\right), \\
R^a & =r.
\end{aligned}
\end{equation}
\blue{The RS ans\"atz is commonly corroborated by numerical simulations on finite-size instances of dynamics. We did not find evidence towards the necessity of breaking the replica symmetry, e.g., considering
one-step replica symmetry breaking~\cite{Mezard-1987}.}

Under this RS ans\"atz, we derive the following free energy functions and associated single-variable effective Hamiltonians for three types of learning
rules. Technical details are given in Appendix~\ref{app-a}.

\textit{Hebbian learning}.---The free energy reads
\begin{equation}
  -\beta f=\frac{1}{2} Q \hat{Q}-q \hat{q} - r \hat{r}-\ln \sigma+k \sqrt{\beta} r q + \int D u D v \ln I,
\end{equation}
where $\sigma \equiv \sqrt{1+g^2 \beta(q-Q)}$, $\hat{Q}, \hat{q}$ and $\hat{r}$ are conjugated order parameters, and 
$I \equiv \int d x e^{\mathcal{H}(x)}$. The effective Hamiltonian of the Hebbian learning is given by
\begin{equation}
  \mathcal{H}(x)\equiv-\beta \eta x^2+\frac{1}{2}(2 \hat{q}-\hat{Q}) \phi^2(x)+\sqrt{\hat{Q}} u \phi(x)- \frac{1}{2\sigma^2}(g \sqrt{\beta Q} v+\hat{r} \phi(x)-\sqrt{\beta} x)^2.
\end{equation}

\textit{Feedback learning}.---The free energy reads
\begin{equation}
  -\beta f=\frac{1}{2} Q \hat{Q}-q \hat{q}-\ln \sigma+\int D u D v \ln I,
\end{equation}
where $I \equiv \int d x e^{\mathcal{H}(x)}$, and the corresponding effective Hamiltonian reads
\begin{equation}
  \mathcal{H}(x) \equiv -\beta \eta x^2+\frac{1}{2}(2 \hat{q}-\hat{Q}) \phi^2(x)+\sqrt{\hat{Q}} u \phi(x)- \frac{1}{2\sigma^2}\left(\sqrt{\beta g^2 Q + \beta \delta^2 q^2} v-\sqrt{\beta} x\right)^2.
\end{equation}

\textit{Homeostatic learning}.---The free energy reads
\begin{equation}
  -\beta f=\frac{1}{2} Q \hat{Q}-q \hat{q} - r \hat{r}-\ln \sigma-k \sqrt{\beta} r q +\int D u D v \ln I,
\end{equation}
where $I \equiv \int d x e^{\mathcal{H}(x)}$, and the corresponding effective Hamiltonian reads
\begin{equation}
  \begin{aligned}   
      \mathcal{H}(x) \equiv-\beta \eta x^2+\frac{1}{2}(2 \hat{q}-\hat{Q}) \phi^2(x)+\sqrt{\hat{Q}} u \phi(x)- \frac{1}{2\sigma^2}(g \sqrt{\beta Q} v+\hat{r} \phi(x)-\sqrt{\beta} x+\sqrt{\beta} k r_{\rm tg} q)^2.
  \end{aligned}
\end{equation}

Next, we send the temperature to zero to concentrate the Boltzmann measure on the ground state, i.e., zero-speed points in the phase space. In this limit,
we have to properly rescale the order parameters because of the divergence behavior observed when $\beta\to\infty$. A reasonable scaling behavior is specified below.
\begin{equation}\label{scaling}
  \begin{aligned}
  & (q-Q) \rightarrow \frac{\chi}{\beta},\\
  & (2 \hat{q}-\hat{Q}) \rightarrow \beta \hat{\chi}, \\
  & \hat{q} \rightarrow \beta^2 \hat{q} ,\\
  & r \rightarrow \sqrt{\beta}r ,\\
  & \hat{r} \rightarrow \sqrt{\beta} \hat{r} ,
  \end{aligned}
\end{equation}
This allows us to obtain the zero-temperature free energy and the corresponding saddle point equations in the thermodynamic limit. The saddle point equations are 
obtained by setting the derivative of the free energy with respect to the associated order parameters and their conjugate counterparts to zero. Detailed analyses are given in Appendix~\ref{app-a}. Here, we summarize the main results.

\textit{Hebbian learning}.---The free energy reads
\begin{equation}
  \begin{aligned}
  -f & =-\frac{1}{2}(q \hat{\chi}+2 \hat{q} \chi)- r \hat{r} + k r q+ \int D u D v \mathcal{H}_0\left(x^*\right),
\end{aligned}
\end{equation}
where $x^*=\operatorname{argmax}_x \mathcal{H}_0(x)$, and
\begin{equation}
  \mathcal{H}_0(x)=- \eta x^2+\frac{1}{2} \hat{\chi} \phi^2(x)+  \sqrt{2 \hat{q} } u \phi(x)- \frac{1}{2\sigma^2}(g \sqrt{q} v+\hat{r} \phi(x)- x)^2,
\end{equation}
with the following zero-temperature saddle-point equations (SDEs):
\begin{equation}\label{hebbian saddle}
  \begin{aligned}
  q & =\left[\phi^2\left(x^*\right)\right] ,\\
  \chi & =\frac{1}{\sqrt{2 \hat{q}}}\left[u \phi\left(x^*\right)\right] ,\\
  \hat{q} & = \frac{g^2}{2\sigma^4}\left(g^2 q+\hat{r}^2\left[\phi^2\left(x^*\right)\right]+\left[\left(x^*\right)^2\right]+2 g \sqrt{q} \hat{r}\left[v \phi\left(x^*\right)\right]-2 g \sqrt{q}\left[v x^*\right]-2 \hat{r}\left[x^* \phi\left(x^*\right)\right]\right) ,\\
  \hat{\chi} & = 2 k r-\frac{g^2}{\sigma^2}-\frac{g \hat{r}}{\sigma^2 \sqrt{q}}\left[v \phi\left(x^*\right)\right]+ \frac{g}{\sigma^2\sqrt{q}}\left[v x^*\right] ,\\
  r & =-\frac{g \sqrt{q}}{\sigma^2}\left[v \phi\left(x^*\right)\right]-\frac{\hat{r}}{\sigma^2}\left[\phi^2\left(x^*\right)\right]+\frac{1}{\sigma^2}\left[x^* \phi\left(x^*\right)\right] ,\\
  \hat{r} & =k q,
  \end{aligned}
\end{equation}
where $\left[ \bullet \right]\equiv\int DuDv \bullet$ as before, and to estimate this average, we first generate $M$ Monte Carlo samples $\{(u_i, v_i)\}_{i = 1}^M$, for each of them we find the global maximum of $\mathcal{H}_0 (x)$, i.e., $x^*$. All these values corresponding to the maxima [for each pair of $(u,v)$]  are further used to complete the calculation of the order parameter for one round of iteration.

\textit{Feedback learning}.---The free energy reads
\begin{equation}
  \begin{aligned}
  -f & =-\frac{1}{2}(q \hat{\chi}+2 \hat{q} \chi) + \int(D u D v) \mathcal{H}_0\left(x^*\right),
\end{aligned}
\end{equation}
where $x^*=\operatorname{argmax}_x \mathcal{H}_0(x)$, and
\begin{equation}
  \mathcal{H}_0(x)=- \eta x^2+\frac{1}{2} \hat{\chi} \phi^2(x)+  \sqrt{2 \hat{q} } u \phi(x)- \frac{1}{2\sigma^2}(\sqrt{ g^2 q +\delta^2 q^2} v- x)^2,
\end{equation}
with the following zero-temperature saddle point equations:
\begin{equation}\label{feedback saddle}
  \begin{aligned}
  q & =\left[\phi^2\left(x^*\right)\right] ,\\
  \chi & =\frac{1}{\sqrt{2 \hat{q}}}\left[u \phi\left(x^*\right)\right] ,\\
  \hat{q} & = \frac{g^2}{2 \sigma^4}\left(g^2 q + \delta^2 q^2+\left[\left(x^*\right)^2\right]-2 \sqrt{g^2 q + \delta^2 q^2}\left[v x^*\right]\right) ,\\
  \hat{\chi} & = -\frac{g^2}{\sigma^2}-\frac{2 q \delta^2}{\sigma^2}+ \frac{g^2 + 2q \delta^2}{\sigma^2\sqrt{g^2 q+\delta^2 q^2}}\left[v x^*\right] ,\\
  r & = -\frac{\sqrt{ g^2 q +\delta^2 q^2}}{\sigma^2}\left[v \phi\left(x^*\right)\right]+\frac{1}{\sigma^2}\left[x^* \phi\left(x^*\right)\right] ,
  \end{aligned}
\end{equation}
where we must remark that the response function in the last line is not a natural order parameter emerging from the disorder average, but can be
computed from a moment generating function detailed in Appendix~\ref{app-a}.

\begin{figure}
    \centering
    \includegraphics[width=1.0\textwidth]{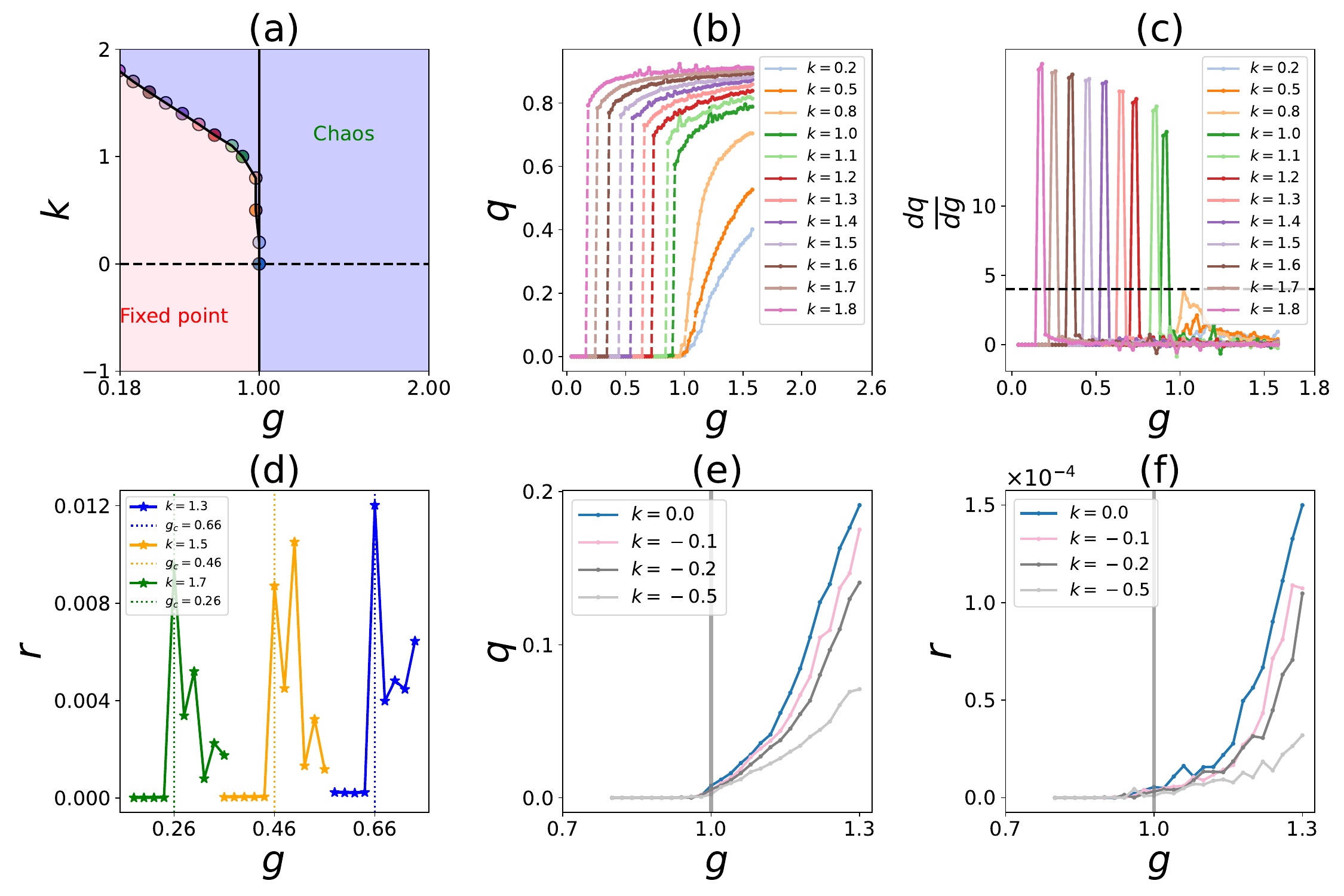}
    \caption{Phase diagram and order parameters with varying $g$ under Hebbian plasticity of different plasticity strengths.
    (a) The phase diagram is divided into two regions: the pink-colored area represents the fixed-point region, while the blue-colored area represents the chaotic region. \blue{The boundary is determined by equating free energies of two phases.}
    (b,d) Plots of order parameters $q$ and $r$ against $g$ for different positive values of $k$. In (c), the plot shows the derivative of $q$ with respect to $g$. 
    The dashed lines in (b) indicate the point where a first-order phase transition occurs, and in (c), the dashed line marks a sharp increase of the order parameter $q$. The threshold for the sharp slope is set to $4.0$. (e-f) Plots of order parameters $q$ and $r$ against $g$ for different negative values of $ k$. The vertical lines in (e,f) indicate the phase transition point. Results are the averages over five independent runs of the SDE solver (see Appendix~\ref{app-b}).}\label{fig1}
\end{figure}

\textit{Homeostatic learning}.---The free energy reads
\begin{equation}
  \begin{aligned}
  -f & =-\frac{1}{2}(q \hat{\chi}+2 \hat{q} \chi)- r \hat{r} -k r q+ \int(D u D v) \mathcal{H}_0\left(x^*\right),
\end{aligned}
\end{equation}
where $x^*=\operatorname{argmax}_x \mathcal{H}_0(x)$, and
\begin{equation}
  \mathcal{H}_0(x)=- \eta x^2+\frac{1}{2} \hat{\chi} \phi^2(x)+  \sqrt{2 \hat{q} } u \phi(x)- \frac{1}{2\sigma^2}(g \sqrt{q} v+\hat{r} \phi(x)- x+ k r_{\rm tg} q)^2,
\end{equation}
with the following zero-temperature SDEs:
\begin{equation}\label{hemo saddle}
  \begin{aligned}
  q & =\left[\phi^2\left(x^*\right)\right] ,\\
  \chi & =\frac{1}{\sqrt{2 \hat{q}}}\left[u \phi\left(x^*\right)\right] ,\\
  \hat{q} & = \frac{g^2}{2\sigma^4}\left(g^2 q+k^2 r_{\rm tg}^2 q^2+\hat{r}^2\left[\phi^2\left(x^*\right)\right]+\left[\left(x^*\right)^2\right]+2 g \sqrt{q} \hat{r}\left[v \phi\left(x^*\right)\right]-2 g \sqrt{q}\left[v x^*\right]-2 \hat{r}\left[x^* \phi\left(x^*\right)\right]\right. \\
  &\left.+2 k \hat{r} r_{\rm tg} q \left[\phi\left(x^*\right)\right] - 2 k r_{\rm tg} q \left[x^*\right]\right) ,\\
  \hat{\chi} & = -2 k r-\frac{g^2}{\sigma^2}-\frac{2 k^2 r_{\rm tg}^2 q}{\sigma^2}-\frac{g \hat{r}}{\sigma^2 \sqrt{q}}\left[v \phi\left(x^*\right)\right]+ \frac{g}{\sigma^2\sqrt{q}}\left[v x^*\right]-\frac{2}{\sigma^2} k \hat{r} r_{\rm tg} \left[\phi\left(x^*\right)\right]+\frac{2}{\sigma^2}  k r_{\rm tg} \left[x^*\right],\\
  r & =-\frac{g \sqrt{q}}{\sigma^2}\left[v \phi\left(x^*\right)\right]-\frac{\hat{r}}{\sigma^2}\left[\phi^2\left(x^*\right)\right]+\frac{1}{\sigma^2}\left[x^* \phi\left(x^*\right)\right]-\frac{kr_{\rm tg}q }{\sigma^2}\left[\phi\left(x^*\right)\right],\\
  \hat{r} & =-k q.\\
  \end{aligned}
\end{equation}

 \begin{figure}
    \centering
    \includegraphics[width=1.0\textwidth]{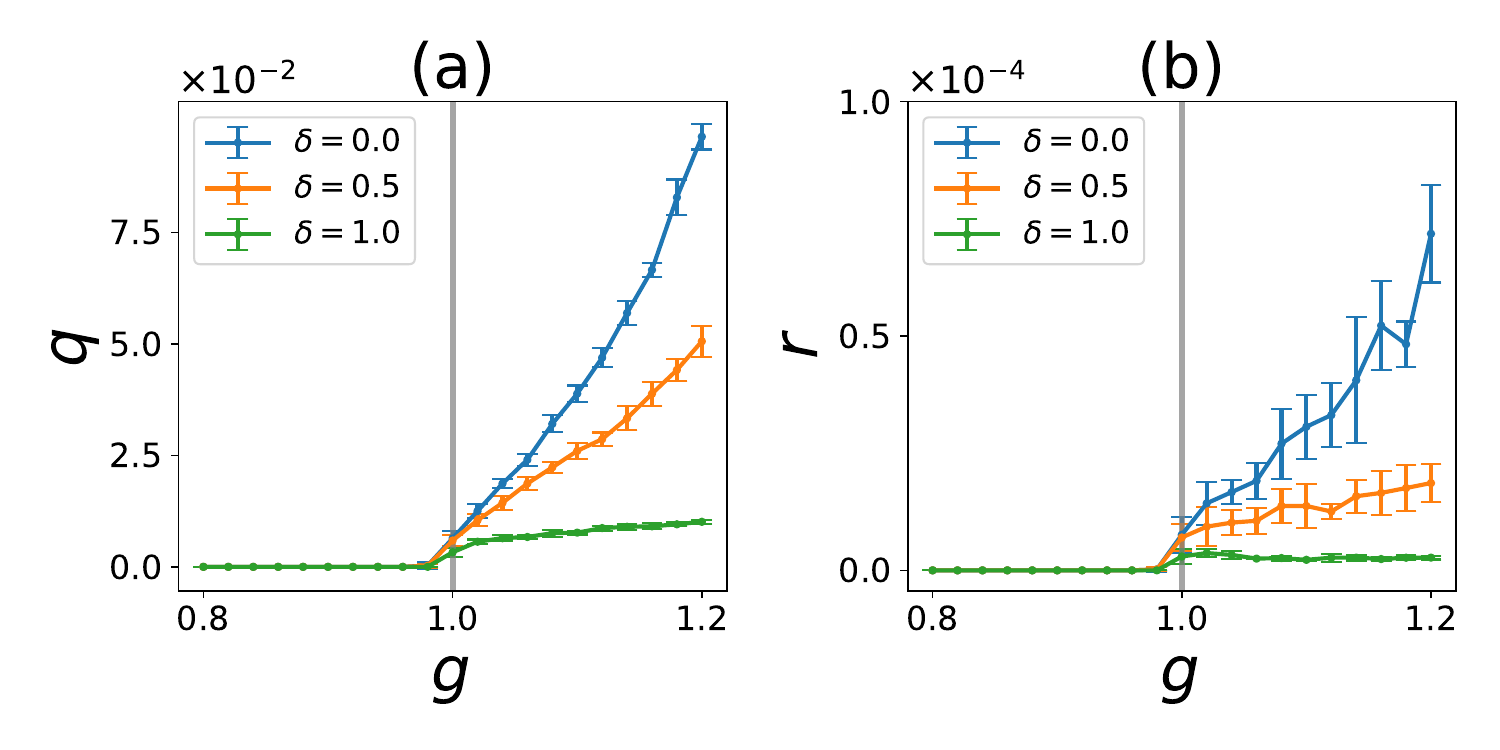}
    \caption{The profile of the order parameters $q$ and $r$ with respect to the gain parameter $g$ and the strength of feedback learning $\delta\in\{0.0, 0.5, 1.0\}$. The vertical lines in (a,b) indicate the phase transition point. Five independent runs of the SDE solver are considered.}\label{fig2}
\end{figure}

\begin{figure}
    \centering
    \includegraphics[width=1.0\textwidth]{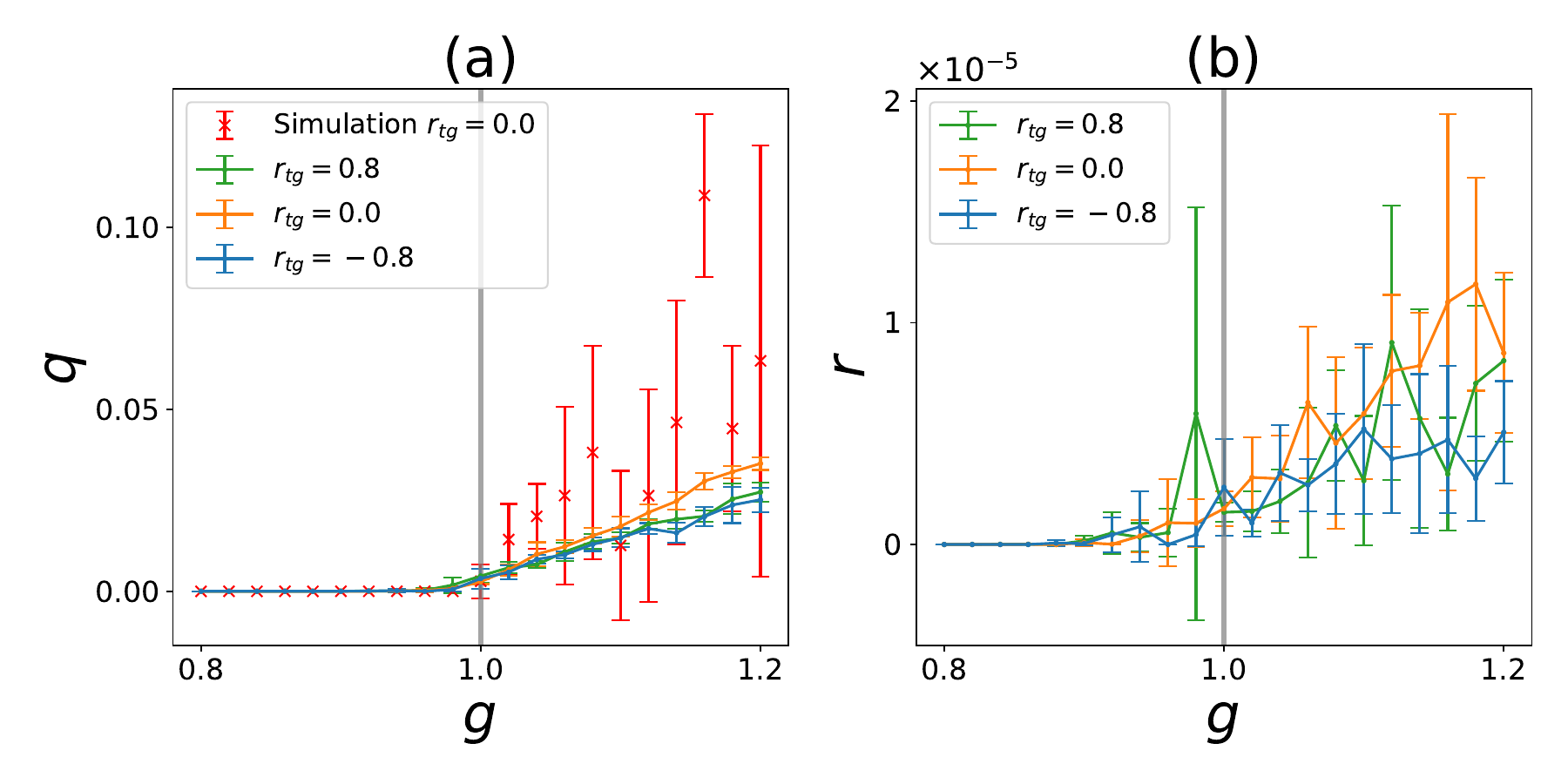}
    \caption{The profile of the order parameters with respect to the gain parameter $g$ and the target firing rate of homeostatic learning $r_{\operatorname{tg}}\in\{-0.8, 0.0, 0.8\}$.
    The learning strength $k$ is set to $0.5$; we also present the simulation results for $ r_{\text{tg}} = 0 $ in networks of $N=1\,000$ ($\tau=0$). The simulation results are averages over the last 
    $500$ time steps of the simulated dynamics. The vertical lines in (a,b) indicate the phase transition point.
    Five independent runs are used to obtain the averaged data points.}\label{fig3}
\end{figure}

\section{Results and discussion}
\subsection{Phase diagram analysis}
For simplicity, we set $\eta=0$ in the following discussion. We first show the theoretical results of Hebbian learning in Fig.~\ref{fig1}. In the $(k,g)$ plane, there appear two different dynamic regimes. When $k<0.8$, the chaos transition is 
continuous, while a larger value of $k$ would trigger a discontinuous chaos transition, as supported by the $q$-profile and its derivatives [see Fig.~\ref{fig1} (b,c)]. In particular, a 
high value of $k$ will shift the onset of a chaos transition to a smaller synaptic gain parameter (than a standard value of $g_c=1$ for $k=0$). Note that in our case, the original non-reciprocal coupling is directly added by the Hebbian term (equivalently, $\tau\to0$---very fast or instantaneous synaptic change). In this sense, our results are different from a recent study of finite $\tau$ using dynamical mean-field theory~\cite{Clark-2024}. Moreover, our current analysis could not distinguish the internal refined structures of chaotic regimes. \blue{Perhaps, additional analysis such as Lyapunov exponents (detailed below) needs to be introduced}. The merit of our method lies in the clarification of dynamical phase transitions through well-defined order parameters optimizing a free energy function, rather than the activity auto-covariance in dynamical mean-field theory, where the transition to chaos is commonly determined by the change of the concavity of a classical potential~\cite{Chaos-1988}. 

 In addition, our theory predicts that the response parameter displays a peak on the onset of the chaos transition point [Fig.~\ref{fig1} (d)]. Moreover, the negative value of $k$ does not alter the transition point and type [Fig.~\ref{fig1} (e,f)]. If a finite speed (but still small in magnitude) of dynamics is considered, i.e., studying the finite temperature case, one would also obtain a peak at the exact onset of the chaos, as already shown in the non-plastic model~\cite{Qiu-2024}.

We next look at the theory of feedback learning (Fig.~\ref{fig2}). Our theory predicts that tuning the feedback strength does not change the type of dynamics transition and the transition
location. However, stronger feedback would limit the dynamics diversity, as expected from the readout error nature of the feedback strength.

\begin{figure}
    \centering
    \includegraphics[width=0.9\textwidth]{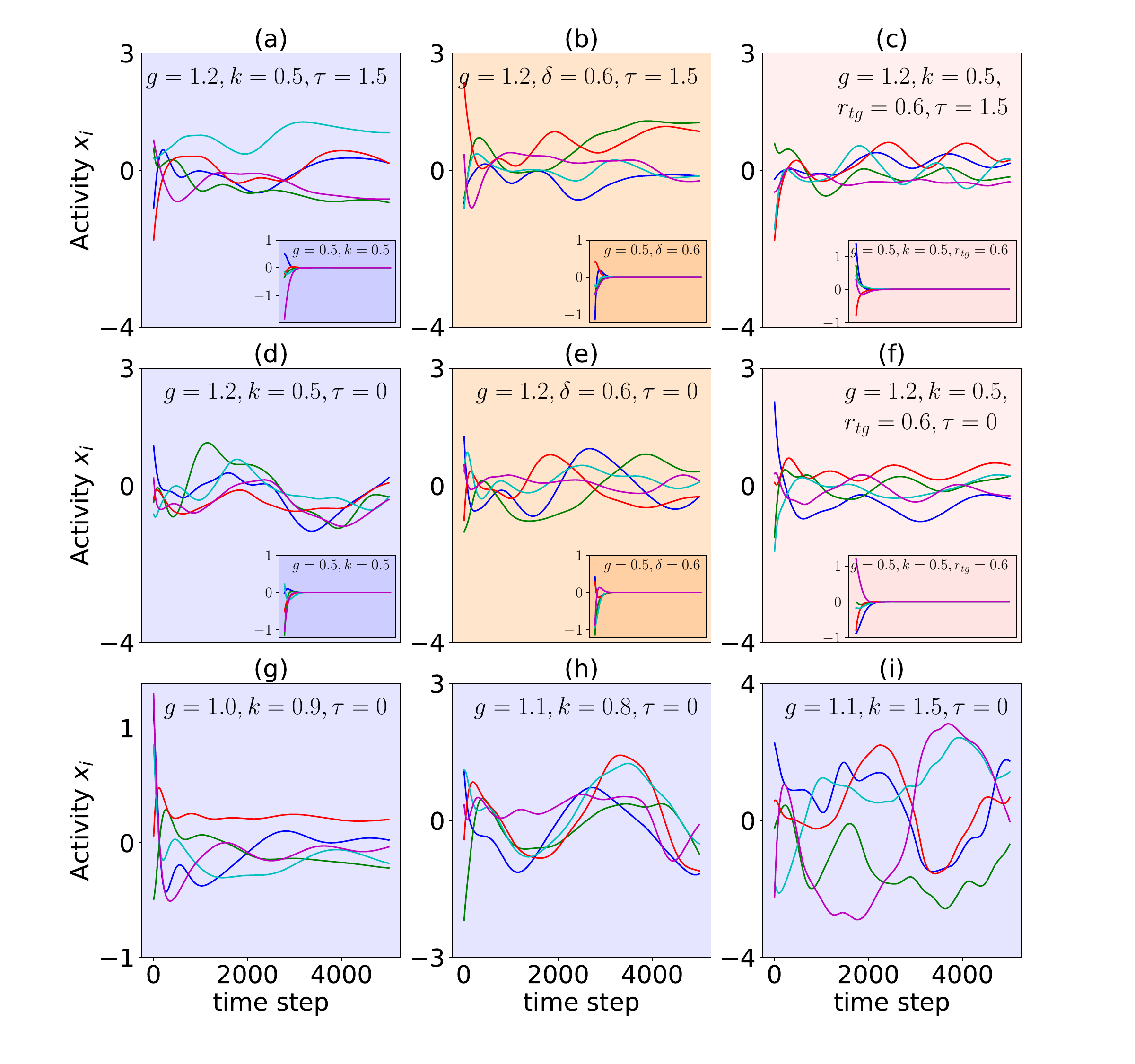}
    \caption{Neural dynamics under different plasticity rules (a network of $1\,000$ neurons, five of which are randomly selected and shown). Three different background colors distinguish different learning types.
    (a, d, g, h, i) Neural dynamics under Hebbian learning. In (a), the time constant $\tau=1.5$; in (d, g, h, i), $\tau=0$. In the main plot, parameters are $(g,k)=(1.2, 0.5)$, and in
     the inset plot parameters are $(g,k)=(0.5, 0.5)$.
    (b, e) Neural dynamics under feedback learning. In (b), $\tau=1.5$; in (e), $\tau=0$. In the main plot, parameters are $(g,\delta)=(1.2, 0.6)$, and in the inset parameters are $(g,\delta)=(0.5, 0.6)$.
    (c, f) Neural dynamics under homeostatic learning. In (c), $\tau=1.5$; in (f), $\tau=0$. In the main plot, parameters are $(g,k)=(1.2, 0.5)$, and $r_{\rm tg} = 0.6 $, and in the inset parameters are $(g,k)=(0.5, 0.5)$, and $ r_{\rm tg} = 0.6 $.}\label{fig4}
\end{figure}

We finally studied the firing rate homeostatic learning. Given the plasticity strength $k$, we do not observe a qualitative change of the chaos transition (including the onset point) when varying different 
target rates (Fig.~\ref{fig3}). In fact, by setting $r_{\rm tg}=0$, we recover the Hebbian learning case with negative strength (or anti-Hebbian plasticity). However, further increasing the target rate will
make the network enter the non-trivial fixed-point phase. We thus conclude that the homeostatic learning does not shift the chaos transition compared to the corresponding non-plastic counterpart, while the activity magnitude can 
be suppressed by tuning the setpoint. Note that in Fig.~\ref{fig3} (a) there exists a gap between the theoretically-predicted unstable-fixed-points activity and the simulations (non-zero speed), as also proved in a recent work~\cite{Wang-2024}.  Figure~\ref{fig4} summarizes the representative dynamics trajectories when different plasticity rules are considered.

\subsection{Lyapunov exponent analysis of combined dynamics}
\blue{The Lyapunov exponent is a fundamental quantity used to characterize chaos in dynamical systems. It measures the exponential rate at which two infinitesimally close trajectories in phase space diverge. A positive maximal Lyapunov exponent (MLE) indicates sensitive dependence on initial conditions---a hallmark of chaotic behavior. In our context, this measure provides a direct way to verify the chaotic nature of the neuron-synapse coupled dynamics.

To evaluate the MLE of our coupled dynamical system, we numerically implement the orbit separation method~\cite{JC-2003,Yu-2025}. We consider two identical copies of the system with an initial perturbation of magnitude $\delta_0=10^{-5}$ introduced to perturb a specific neuronal state. The dynamical equations (governing both neuronal activity and synaptic variables) are evolved simultaneously for both trajectories. At each time step $t$, we compute the separation $\delta(t)$ in the phase space. The separation is regularly renormalized to
 maintain a small separation after every update of the trajectories, and the MLE is estimated as:
\begin{equation}
\lambda_{\max }=\lim _{T \rightarrow \infty} \frac{1}{T} \sum_{t=1}^T \ln\left( \frac{\delta(t)}{\delta_0}\right),
\end{equation}
where $\delta(t)$ is the distance between the two trajectories after one update at the $t$-th iteration, and the sum is averaged over long simulation time after transient relaxation.}

\begin{figure}
    \centering
    \includegraphics[width=1.0\textwidth]{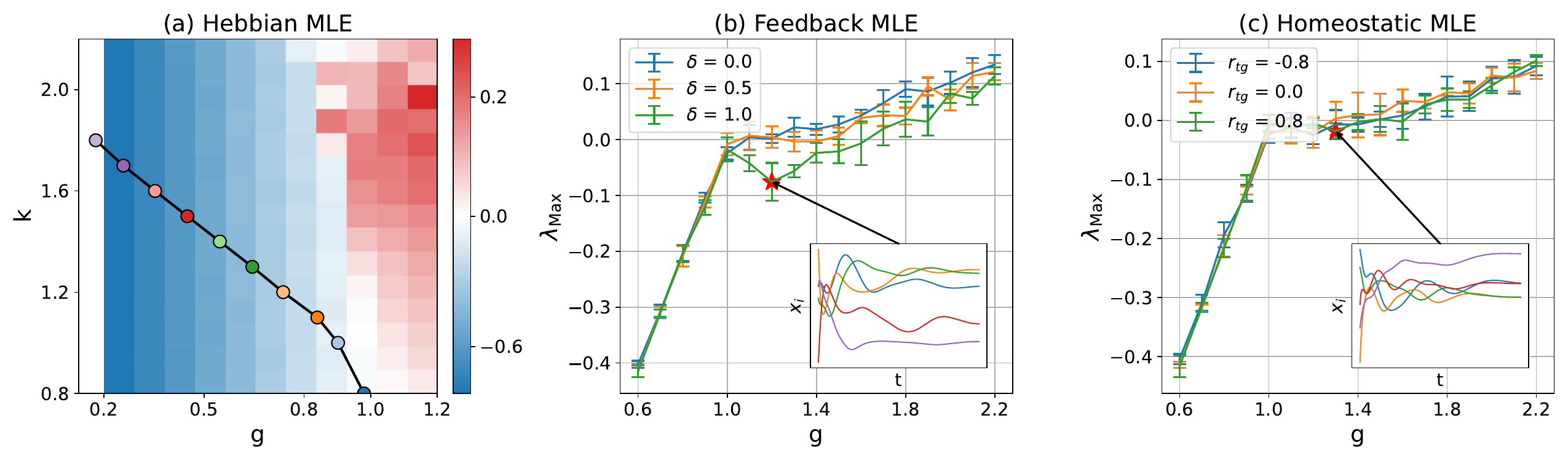}
    \caption{\blue{Maximal Lyapunov exponent $\lambda_{\max }$ as a function of plasticity strength $g$ for different types of synaptic plasticity, computed by orbit separation method with N = 1000 and an initial perturbation of magnitude $\delta_0=10^{-5}$. The exponents are averaged over five independent estimates. (a) Hebbian learning. The connected symbols indicate a chaos transition boundary (continuous or discontinuous, depending on the value of $k$). (b) Feedback learning. (c) Homeostatic learning ($k=0.5$). Insets display typical examples of dynamics trajectries.} }\label{fig5}
\end{figure}

\blue{We evaluate $\lambda_{\text {max }}$ for each of the three synaptic plasticity rules introduced in our study across a range of gain strength $g$ and relevant plasticity parameters (e.g., Hebbian coupling strength $k$, feedback strength $\delta$, homeostatic strength $r_{\rm tg}$ ). As the Hebbian coupling strength $k$ increases, the phase transition point for the onset of chaos originally located at $g_c=1$, moves leftward [Fig.~\ref{fig5} (a)]. The boundary in the right bottom part is approximately consistent with the MLE estimates [Fig.~\ref{fig5} (a)], i.e., a transition from negative to positive MLE. However, in the left part of Fig.~\ref{fig5} (a), even at or above the transition line (upper-left regime), we observe a negative maximal Lyapunov exponent. This discrepancy arises because, within this parameter region, numerous stable non-zero fixed points coexist with (transient) chaotic states where the order parameter $q$ remains positive. This feature was also observed in a slow-synapse model~\cite{Clark-2024}.

For the case of feedback learning [Fig.~\ref{fig5} (b)], 
the MLE increases monotonically with $g$ but the overall magnitude decreases with increasing $\delta$. Notably, around $g=1.2$ for a value of $\delta=1$, the MLE becomes negative, indicating the presence of non‑trivial fixed points rather than chaotic attractors (see the inset).
The feedback strength does not alter the location of the phase transition point, which aligns well with MLE estimates.

For the case of homeostatic learning [Fig.~\ref{fig5} (c)], 
the MLE increases with $g$ and the phase transition point coincides with the prediction of the replica theory. The inset demonstrates an example where the dynamics reach a non‑trivial fixed point in this regime. Additionally, variations in the parameter $r_{\rm tg}$ have minimal influence on the overall dynamics in this regime.}

\subsection{Learning performances in desired-output tasks}
\begin{figure}
    \centering
    \includegraphics[width=1.0\textwidth]{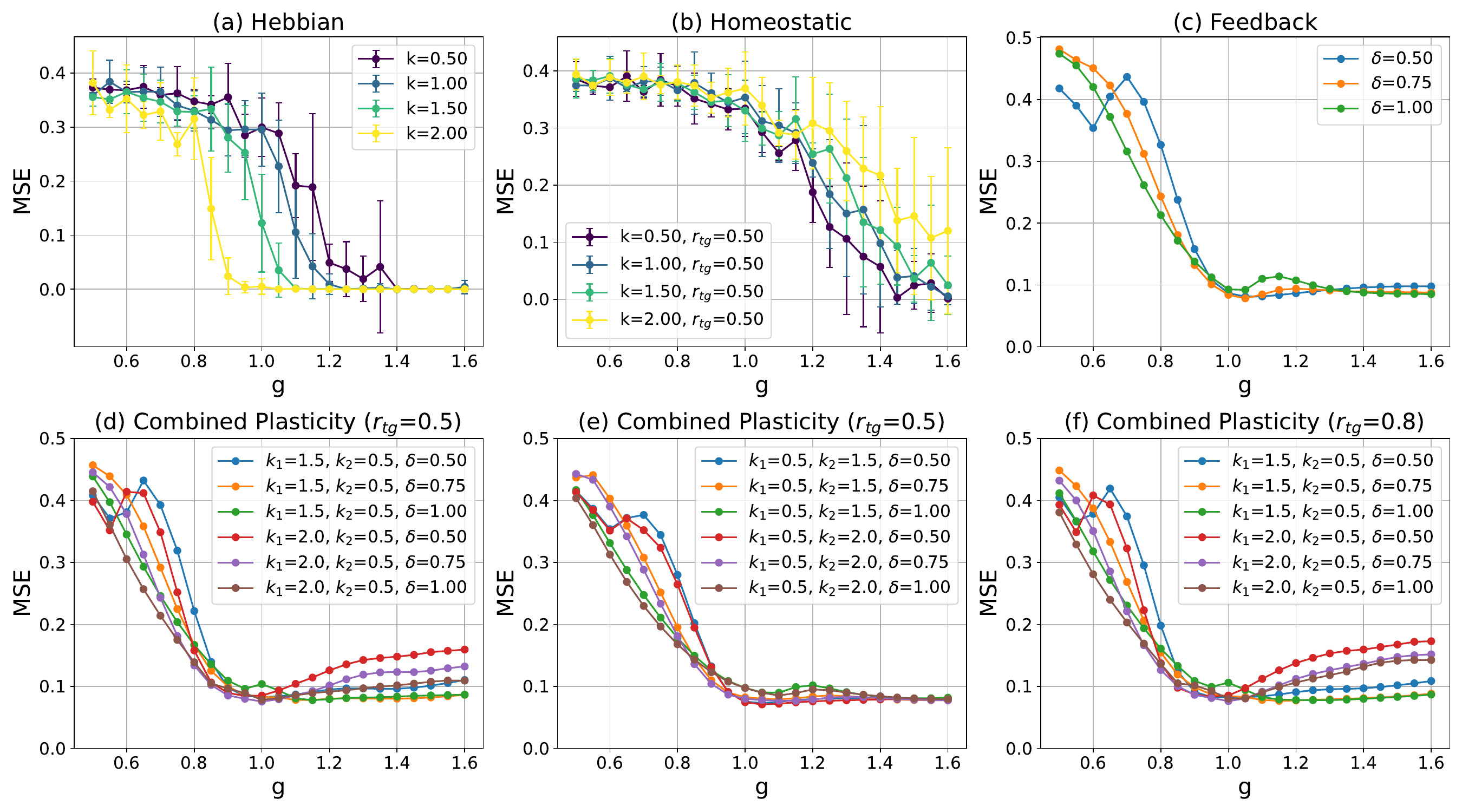}
    \caption{
\blue{Reconstruction performance curves for different network dynamics under various plasticity rules ($N=1000$). The target signal is $z(t)=\sin(\omega t)$ with $\omega=0.02\pi$. Performance is quantified by the mean squared error (MSE) between the network output $\hat{z}(t)$ and the target signal $z(t)$.  
(a) Hebbian dynamics. Results are averaged over five random initializations using the optimal linear readout~\cite{HWZ-2025}.  
(b) Homeostatic dynamics. Results are averaged over five random initializations using the optimal linear readout. Error bars denote variability across independent trials. 
(c) Feedback learning. FORCE learning algorithm is used for the readout training~\cite{Abbott-2012}. 
(d-f) Combined-plasticity-driven dynamics.} }\label{fig6}
\end{figure}

\begin{figure}
    \centering
    \includegraphics[width=1.0\textwidth]{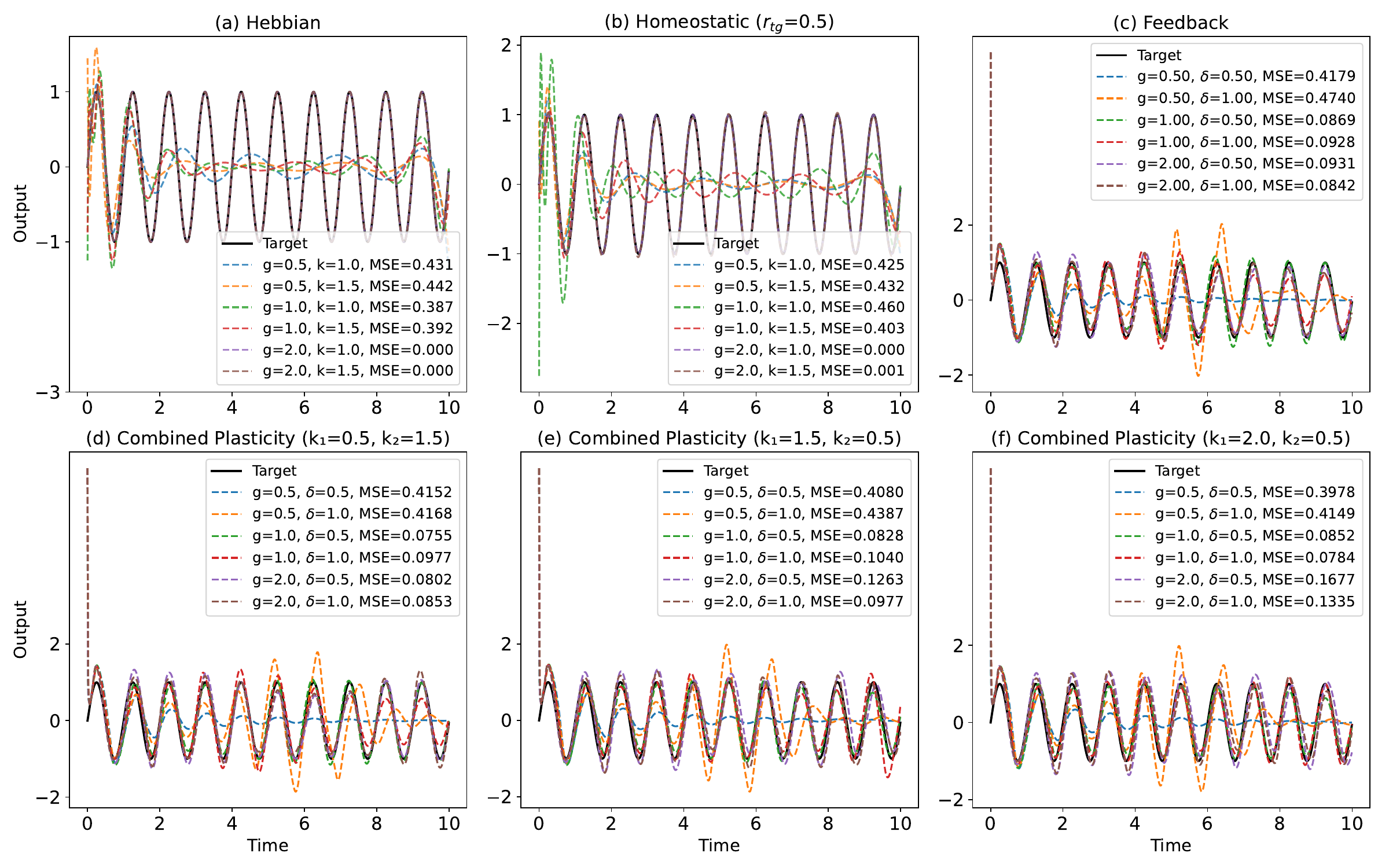}
    \caption{
\blue{Trajectory comparison between the network output $\hat{z}(t)$ and the target signal $z(t)$ over $1000$ time steps for representative parameter settings of different plasticity dynamics. The target signal is $z(t)=\sin(\omega t)$ with $\omega=0.02\pi$. The selected parameter values and the corresponding mean squared errors (MSE) are indicated in the figure legend. Results show the reconstruction quality achieved by Hebbian, feedback, homeostatic, and combined plasticity mechanisms under the same description as in Fig.~\ref{fig6}.}
    }\label{fig7}
\end{figure}

\blue{To quantitatively evaluate the computational advantage of the separate or combined plasticity mechanisms, we focus on a reservoir computing framework in which the recurrent network serves as a nonlinear dynamical reservoir, and only a linear readout is trained to decode the network state for a specific computational task, e.g., we consider a task of desired output. Specifically, the network is required to output a target sinusoidal signal based on the reservoir neural activity. This setting provides a principled measure of the computational power endowed by different plasticity-driven dynamics.
The performance is quantified by the mean squared error (MSE) between the predicted output of the optimally trained linear readout and the target sinusoidal signal,
\begin{equation}
\epsilon = \left\langle \left( z(t) - \hat{z}(t) \right)^2 \right\rangle,
\end{equation}
where $\langle\cdot\rangle$ denotes a temporal expectation, $\hat{z}(t)$ denotes the output of the trained linear readout, and $z(t)$ is the target signal.

In the feedback learning, the network dynamics evolve according to the following discretization:
\begin{equation}
\mathbf{x}(t+\Delta t)
=
\mathbf{x}(t)
+
\Delta t \left[
-\mathbf{x}(t)
+ \mathbf{J}\,\tanh(\mathbf{x}(t))
+ \mathbf{u}\, z(t)
\right],
\end{equation}
where $\mathbf{J}$ is a fixed random recurrent matrix and $\mathbf{u}$ is the feedback vector defined as
\begin{equation}
\mathbf{u} = \delta \cdot \boldsymbol{\xi},
\qquad
\boldsymbol{\xi} \sim \mathcal{N}(0,\mathbf{I}).
\end{equation}
The network output is given by
\begin{equation}
z(t) = \mathbf{w}^\top \tanh(\mathbf{x}(t)),
\end{equation}
where $\mathbf{w}$ adapts via the FORCE algorithm (see Appendix~\ref{app-c} for details). 

For the combined plasticity model, we introduce a linear combination of Hebbian and homeostatic terms to the baseline recurrent connectivity:
\begin{equation}
\mathbf{W}(t) =
\mathbf{J}
+
\frac{k_1-k_2}{N}\tanh[\bx(t)]\tanh[\bx(t)]^\top+\frac{k_2}{N}\mathbf{r}_{\rm tg}\tanh[\bx(t)]^\top,
\end{equation}
where $\mathbf{r}_\mathrm{tg}$ is a constant vector with elements equal to $r_\mathrm{tg}$, and $k_1$ and $k_2$ denote the Hebbian strength and homeostatic strength, respectively. The dynamics of the combined model are then given by
\begin{equation}
\mathbf{x}(t+\Delta t)
=
\mathbf{x}(t)
+
\Delta t \left[
-\mathbf{x}(t)
+ \mathbf{W}(t)\,\tanh(\mathbf{x}(t))
+ \mathbf{u}\, z(t)
\right].
\end{equation}

In Fig.~\ref{fig6}, we compare the reconstruction performance of networks endowed with
Hebbian, feedback, homeostatic, and combined plasticity dynamics across
different parameter regimes. For the Hebbian learning, increasing the coupling
strength $k$ significantly improves the reconstruction accuracy, and the critical value of synaptic gain becomes smaller, while in the
homeostatic case, for a fixed $r_{\rm tg}$, stronger values of $k$ degrade
the performance, especially at a large value of $g$. For the feedback learning, increasing the feedback strength
$\delta$ yields a modest improvement in reconstruction. The critical region for the performance transition shifts to the right-hand side of the edge of chaos predicted by theory, probably due to the finite-value effect of the network size. The phase of non-trivial fixed points around the discontinuous transition may not be computationally useful for this task.

The performance of the combined-plasticity scenario is shown in Fig.~\ref{fig6} (d-f). The performance curves resemble those of feedback learning but achieve slightly better accuracy overall. However, compared with pure Hebbian
learning, a performance plateau emerges at $g>1$. In particular, the parameter setting of $k_1<k_2$ maintains a relatively lower and identical plateau.
These results indicate that, for this specific reconstruction task, the combined plasticity mechanism is not optimal but displays a robust best performance at $g=1$ (this is the continuous onset of chaos for homeostatic and feedback learning). In Fig.~\ref{fig7}, we further illustrate this behavior by comparing the trajectories of $\hat{z}(t)$ and $z(t)$ for representative parameter choices corresponding to those analyzed in Fig.~\ref{fig6}.}

\section{Concluding remarks}
Coupled dynamics between neurons and synapses are ubiquitous in the brain. Understanding the nature of neuron-synapse interaction remains challenging in both theory and experiments. Here, we apply the quasi-potential method developed recently in analyzing non-plastic recurrent networks to address how learning induces dynamical transitions in recurrent neural networks. Three types of plasticity rules are considered: Hebbian, feedback, and homeostatic plasticities. For simplification, we do not take into account the intrinsic time scales of synaptic dynamics, but instead, we focus on the fixed-point structure of the dynamics phase space. Our theoretical calculation reveals that in Hebbian learning, the Hebbian strength will alter the nature of the chaos transition. More precisely, when the strength grows above some threshold, a discontinuous chaos transition occurs at a smaller synaptic gain compared to the non-plastic network. Decreasing the Hebbian strength, a continuous chaos transition will be recovered. The dynamics simulation \blue{and Lyapunov exponent analysis} support this theoretical finding. However, the feedback and homeostatic learning do not alter the location and type of the chaos transition. Only the chaotic fluctuation is tuned by the feedback strength, while a large target setpoint suppresses
the chaotic fluctuation in the case of homeostatic learning. \blue{Roles of different plasticity rules are also explored in a desired-output computational task using reservoir computing, consistent with the theoretical analysis.}

\blue{In real brain circuits, neurons are not fully-connected, and the baseline synaptic strengths are not perfectly Gaussian random values. In addition, we consider only the instantaneous synaptic change (time scales of synaptic dynamics are overlooked), which is not biologically plausible. Furthermore, the emergence of chaotic behavior stemming from the profile of the neuronal response latency (neuronal plasticity) was experimentally demonstrated~\cite{EPL-2014}, and neuronal response failures will slow down the synaptic plasticity~\cite{FNC-2015,EPL-2017}. These biological features should be considered in further extension of our works.}

 In future works, it would also be interesting to study the coupled dynamics by tuning separate inverse temperatures and monitoring the competition between neural and synaptic dynamics. It is also promising to further investigate the impact of the change of chaos transition in improving generalization performances in recurrent computation~\cite{EoC-1990,EoC-2004,Maass-2009}, and even verify our theoretical predictions in neurobiological experiments. The study of coupled dynamical systems may also yield insights into neurological and psychiatric disorders.

\section*{Acknowledgments}
 This research was supported by the National Natural Science Foundation of China for
Grant numbers 12475045 and 12122515, and Guangdong Provincial Key Laboratory of Magnetoelectric Physics and Devices (No. 2022B1212010008), and Guangdong Basic and Applied Basic Research Foundation (Grant No. 2023B1515040023).
\section*{Code availability}
Codes to reproduce all results are deposited in our Github~\cite{code-2025}.

\onecolumngrid
\appendix
\section{Replica analysis details of non-equilibrium learning dynamics}\label{app-a}
\subsection{Hebbian learning}\label{app-a1}
Starting from the replicated partition function in the main text, we perform the disorder average as follows,
\begin{equation}\label{Heb1}
  \begin{aligned}
      \left\langle Z_\ell^n  \right\rangle & = \left\langle \int d \mathbf{x} \exp \left[-\beta\left(\frac{1}{2} \sum_{i a}\left(-x_i^a+\sum_j J_{i j} \phi\left(x_j^a\right)\right)^2+\eta \sum_a\left\|\mathbf{x}^a\right\|^2\right)\right] \right\rangle\\
  & = \left\langle \int d \mathbf{x} D \hat{\mathbf{x}} \exp \left[i \sqrt{\beta} \sum_{i a} \hat{x}_i^a\left(-x_i^a+\sum_j J_{i j} \phi\left(x_j^a\right)\right)-\beta \eta \sum_a\left\|\mathbf{x}^a\right\|^2\right] \right\rangle.
  \end{aligned}
\end{equation}
Due to the i.i.d. nature of the coupling statistics, the disorder average is straightforward.
\begin{equation}
\begin{aligned}
& \left\langle\exp \left[i \sqrt{\beta} \sum_{i j} J_{i j} \sum_a \hat{x}_i^a \phi\left(x_j^a\right)\right]\right\rangle \\
= & \left\langle\exp \left[i \sqrt{\beta} \sum_{i j} J_{i j}^0 \sum_a \hat{x}_i^a \phi\left(x_j^a\right)\right]\right\rangle \exp \left[i \frac{k \sqrt{\beta}}{N} \sum_{i j} \sum_a \hat{x}_i^a \phi\left(x_i^a\right) \phi^2\left(x_j^a\right)\right] \\
= & \exp \left[-\frac{1}{2} \frac{\beta g^2}{N} \sum_{i j}\left(\sum_a \hat{x}_i^a \phi\left(x_j^a\right)\right)^2+i \frac{k \sqrt{\beta}}{N} \sum_{i j} \sum_a \hat{x}_i^a \phi\left(x_i^a\right) \phi^2\left(x_j^a\right)\right] \\
= & \exp \left[-\frac{1}{2} \frac{\beta g^2}{N} \sum_{i j} \sum_{a b} \hat{x}_i^a \hat{x}_i^b \phi\left(x_j^a\right) \phi\left(x_j^b\right)+i \frac{k \sqrt{\beta}}{N} \sum_{i j} \sum_a \hat{x}_i^a \phi\left(x_i^a\right) \phi^2\left(x_j^a\right)\right] \\
= & \exp \left[-\frac{1}{2} \beta g^2 \sum_i \sum_{a b} \hat{x}_i^a \hat{x}_i^b Q^{a b}+i k \sqrt{\beta} N \sum_a R^a Q^{a a}\right],
\end{aligned}
\end{equation}
where we have defined two kinds of order parameters:
\begin{equation}
\begin{aligned}
Q^{a b} & =\frac{1}{N} \sum_i \phi\left(x_i^a\right) \phi\left(x_i^b\right), \\
R^a & =\frac{1}{N} \sum_i \hat{x}_i^a \phi\left(x_i^a\right).
\end{aligned}
\end{equation}

Inserting the following identity into Eq.~\eqref{Heb1}, we can further simplify the above result.
\begin{equation}
\begin{aligned}
1 & =\prod_{a \leq b} \int d Q^{a b} \delta\left(Q^{a b}-\frac{1}{N} \sum_i \phi\left(x_i^a\right) \phi\left(x_i^b\right)\right) \prod_a \int d R^a \delta\left(R^a-\frac{1}{N} \sum_i \hat{x}_i^a \phi\left(x_i^a\right)\right) \\
& =\int \frac{d \mathbf{Q} d \hat{\mathbf{Q}} d \mathbf{R} d \hat{\mathbf{R}}}{2 \pi} \exp \left[-i \sum_{a \leq b} Q^{a b} \hat{Q}^{a b}-i \sum_a R^a \hat{R}^a+i \frac{1}{N} \sum_i \sum_{a \leq b} \hat{Q}^{a b} \phi\left(x_i^a\right) \phi\left(x_i^b\right)\right. \\
& \left.+i \frac{1}{N} \sum_a \hat{R}^a \sum_i \hat{x}_i^a \phi\left(x_i^a\right)\right] \\
& =\int \frac{d \mathbf{Q} d \hat{\mathbf{Q}} d \mathbf{R} d \hat{\mathbf{R}}}{2 \pi i / N} \exp \left[-N \sum_{a \leq b} Q^{a b} \hat{Q}^{a b}-N \sum_a R^a \hat{R}^a+\sum_i \sum_{a \leq b} \hat{Q}^{a b} \phi\left(x_i^a\right) \phi\left(x_i^b\right) \right.\\
&\left.+i \sum_a \hat{R}^a \sum_i \hat{x}_i^a \phi\left(x_i^a\right)\right],
\end{aligned}
\end{equation}
where we have rescaled the order parameters as $\hat{Q}^{a b} \rightarrow-i N \hat{Q}^{a b}, R^a \rightarrow-i R^a$ and $\hat{R}^a \rightarrow N \hat{R}^a$.
Therefore, the averaged replicated partition function becomes
\begin{equation}
\begin{aligned}
\left\langle Z_\ell^n\right\rangle & \propto  \int d \mathbf{x} D \hat{\mathbf{x}} d \mathbf{Q} d \hat{\mathbf{Q}} d \mathbf{R} d \hat{\mathbf{R}} \exp \left[-i \sqrt{\beta} \sum_{i a} x_i^a \hat{x}_i^a-\beta \eta \sum_{i a}\left(x_i^a\right)^2-\frac{1}{2} g^2 \beta \sum_{a b} Q^{a b}\left(\sum_i \hat{x}_i^a \hat{x}_i^b\right)\right. \\
& \left.+k \sqrt{\beta} N \sum_a R^a Q^{a a}-N \sum_{a \leq b} Q^{a b} \hat{Q}^{a b}-N \sum_a R^a \hat{R}^a+\sum_i \sum_{a \leq b} \hat{Q}^{a b} \phi\left(x_i^a\right) \phi\left(x_i^b\right) \right. \\
& \left.+i \sum_a \hat{R}^a \sum_i \hat{x}_i^a \phi\left(x_i^a\right)\right] \\
&= \int d \mathbf{Q} d \hat{\mathbf{Q}} d \mathbf{R} d \hat{\mathbf{R}} \exp \left[N\left(-\sum_{a \leq b} Q^{a b} \hat{Q}^{a b}-\sum_a R^a \hat{R}^a+G\right)\right],
\end{aligned}
\end{equation}
where $ \propto$ means that we have neglected irrelevant pre-factors, and the auxiliary quantity
\begin{equation}\label{EqG}
\begin{aligned}
G= & \ln \int d \mathbf{x} D \hat{\bx} \exp \left[-i \sqrt{\beta} \sum_a x^a \hat{x}^a-\beta \eta \sum_a\left(x^a\right)^2-\frac{1}{2} g^2 \beta \sum_{a b} Q^{a b} \hat{x}^a \hat{x}^b+\sum_{a \leq b} \hat{Q}^{a b} \phi\left(x^a\right) \phi\left(x^b\right)\right. \\
& \left.+k \sqrt{\beta} \sum_a R^a Q^{a a}+i \sum_a \hat{R}^a \hat{x}^a \phi\left(x^a\right)\right].
\end{aligned}
\end{equation}

To proceed, we have to adopt the RS ans\"atz:
\begin{equation}
\begin{aligned}
Q^{a b} & =q \delta_{a b}+Q\left(1-\delta_{a b}\right), \\
R^a & =r.
\end{aligned}
\end{equation}
It then follows that
\begin{equation}
\begin{aligned}
& -\frac{1}{2} g^2 \beta \sum_{a b} Q^{a b} \hat{x}^a \hat{x}^b+\sum_{a \leq b} \hat{Q}^{a b} \phi\left(x^a\right) \phi\left(x^b\right) \\
= & -\frac{1}{2} g^2 \beta\left(Q\left(\sum_a \hat{x}^a\right)^2+(q-Q) \sum_a\left(\hat{x}^a\right)^2\right)+\frac{1}{2} \hat{Q}\left(\sum_a \phi\left(x^a\right)\right)^2+\left(\hat{q}-\frac{1}{2} \hat{Q}\right)\left(\sum_a \phi^2\left(x^a\right)\right).
\end{aligned}
\end{equation}
We can then observe that the $n$-dimensional integral in Eq.~\eqref{EqG} factorizes over the replica index.

After completing the following integral over $\hat{x}$:
\begin{equation}
\begin{aligned}
& \int D \hat{x} \exp \left[-\frac{1}{2} g^2 \beta(q-Q) \hat{x}^2+i(g \sqrt{\beta Q} v+\hat{r} \phi(x)-\sqrt{\beta} x) \hat{x}\right] \\
= & \frac{1}{\sigma} \exp \left[-\frac{1}{2} \frac{1}{\sigma^2}(g \sqrt{\beta Q} v+\hat{r} \phi(x)-\sqrt{\beta} x)^2\right],
\end{aligned}
\end{equation}
where $\sigma \equiv \sqrt{1+g^2 \beta(q-Q)}$, we arrive at the following neat formula:
\begin{equation}
G=-n \ln \sigma+k \sqrt{\beta} n r q+\ln \int D u D v I^n ,\\
\end{equation}
where $I \equiv \int d x e^{\mathcal{H}(x)}$ and the effective single-variable Hamitonian can be read off,
\begin{equation}
    \mathcal{H}(x) \equiv-\beta \eta x^2+\frac{1}{2}(2 \hat{q}-\hat{Q}) \phi^2(x)+\sqrt{\hat{Q}} u \phi(x)- \frac{1}{2\sigma^2}(g \sqrt{\beta Q} v+\hat{r} \phi(x)-\sqrt{\beta} x)^2.
\end{equation}
Following the replica trick, the free energy in finite temperature is given by
\begin{equation}
-\beta f=\frac{1}{2} Q \hat{Q}-q \hat{q} - r \hat{r}-\ln \sigma+k \sqrt{\beta} r q + \int D u D v \ln I.
\end{equation}

However, we are interested in the zero temperature limit, which makes us see the fixed points of the non-gradient out-of-equilibrium dynamics. From the scaling
behavior of the order parameters in the main text, one can derive that
\begin{equation}
    \begin{aligned}
    & Q=q-(q-Q) \Rightarrow Q=q-\frac{1}{\beta} \chi \rightarrow  q ,\\
    & \hat{Q}=2 \hat{q}-(2 \hat{q}-\hat{Q}) \Rightarrow \hat{Q}=2 \beta^2 \hat{q}-\beta \hat{\chi} \rightarrow 2 \beta^2 \hat{q}.
    \end{aligned}
\end{equation}
Hence, the effective Hamiltonian $\mathcal{H}(x)$ behaves as an explicit linear function of $\beta$, written as follows.
\begin{equation}\label{Heb0}
    \begin{aligned}
    \mathcal{H}(x) & \equiv-\beta \eta x^2+\frac{1}{2}(2 \hat{q}-\hat{Q}) \phi^2(x)+\sqrt{\hat{Q}} u \phi(x)- \frac{1}{2\sigma^2}(g \sqrt{\beta Q} v+\hat{r} \phi(x)-\sqrt{\beta} x)^2 \\
    & \Rightarrow -\beta \eta x^2+\frac{1}{2}\beta \hat{\chi} \phi^2(x)+ \beta \sqrt{2 \hat{q} } u \phi(x)- \frac{\beta}{2\sigma^2}(g \sqrt{q} v+\hat{r} \phi(x)- x)^2 \equiv \beta \mathcal{H}_0(x).
    \end{aligned}
\end{equation}
Here $\sigma=\sqrt{1+\beta g^2(q-Q)} \rightarrow \sqrt{1+g^2 \chi}$ is a quantity of $\mathcal{O}(1)$. This result further implies that,
\begin{equation}
    \begin{aligned}
    -\beta f & =\frac{1}{2} Q \hat{Q}-q \hat{q} - r \hat{r}-\ln \sigma+k \sqrt{\beta} r q+\int D u D v \ln \int d x e^{\mathcal{H}(x)} \\
    & =-\frac{1}{2}\left[q(2 \hat{q}-\hat{Q})+\hat{Q}(q-Q)\right] - r \hat{r}-\ln \sigma+k \sqrt{\beta} r q+\int D u D v \ln \int d x e^{\mathcal{H}(x)} ,\\
    \Rightarrow- \beta f & \rightarrow -\frac{1}{2} \beta(q \hat{\chi}+2 \hat{q} \chi) - \beta r \hat{r}-\ln \sigma + \beta k r q+\int D u D v \ln \int d x e^{\beta \mathcal{H}_0(x)}.
\end{aligned}
\end{equation}

We can then conclude that the free energy in zero temperature must take the following form:
\begin{equation}\label{Hebf0}
 -f =-\frac{1}{2}(q \hat{\chi}+2 \hat{q} \chi)- r \hat{r} + k r q+ \int D u D v \mathcal{H}_0\left(x^*\right),
 \end{equation}
where $x^*=\operatorname{argmax}_x \mathcal{H}_0(x)$. $\mathcal{H}_0(x)$ has been derived in Eq.~\eqref{Heb0}.
\begin{equation}
    \mathcal{H}_0(x)=- \eta x^2+\frac{1}{2} \hat{\chi} \phi^2(x)+  \sqrt{2 \hat{q} } u \phi(x)- \frac{1}{2\sigma^2}(g \sqrt{q} v+\hat{r} \phi(x)- x)^2.
\end{equation}
Optimization of $\mathcal{H}_0(x)$ is a direct consequence of applying Laplace method to obtain the last integral in Eq.~\eqref{Hebf0}.

Finally, we can set the derivative of the zero-temperature free energy with respect to all relevant order parameters to zero, from which the SDEs can be derived as follows.
\begin{equation}
    \begin{aligned}
    q & =\left[\phi^2\left(x^*\right)\right],\\
    \chi & =\frac{1}{\sqrt{2 \hat{q}}}\left[u \phi\left(x^*\right)\right] ,\\
    \hat{q} & = \frac{g^2}{2\sigma^4}\left(g^2 q+\hat{r}^2\left[\phi^2\left(x^*\right)\right]+\left[\left(x^*\right)^2\right]+2 g \sqrt{q} \hat{r}\left[v \phi\left(x^*\right)\right]-2 g \sqrt{q}\left[v x^*\right]-2 \hat{r}\left[x^* \phi\left(x^*\right)\right]\right) ,\\
    \hat{\chi} & = 2 k r-\frac{g^2}{\sigma^2}-\frac{g \hat{r}}{\sigma^2 \sqrt{q}}\left[v \phi\left(x^*\right)\right]+ \frac{g}{\sigma^2\sqrt{q}}\left[v x^*\right] ,\\
    r & =-\frac{g \sqrt{q}}{\sigma^2}\left[v \phi\left(x^*\right)\right]-\frac{\hat{r}}{\sigma^2}\left[\phi^2\left(x^*\right)\right]+\frac{1}{\sigma^2}\left[x^* \phi\left(x^*\right)\right] ,\\
    \hat{r} & =k q,
    \end{aligned}
\end{equation}
where $\left[ \bullet \right]\equiv\int DuDv \bullet$ as before, and to estimate this average, we first generate $M$ Monte Carlo samples $\{(u_i, v_i)\}_{i = 1}^M$, for each of them we find the global maximum of $\mathcal{H}_0 (x)$, i.e., $x^*$. All these values corresponding to the maxima [for each pair of $(u,v)$]  are further used to complete the calculation of the order parameter for one round of iteration.

\subsection{Feedback learning}
We next turn to the feedback learning. Note that in the feedback learning, we have an extra randomness from the feedback weight $\mathbf{u}$. The disorder average part is thus calculated as follows.
 \begin{equation}
    \begin{aligned}
    & \left\langle\exp \left[i \sqrt{\beta} \sum_{i j} J_{i j} \sum_a \hat{x}_i^a \phi\left(x_j^a\right)\right]\right\rangle \\
    = & \left\langle\exp \left[i \sqrt{\beta} \sum_{i j} J_{i j}^0 \sum_a \hat{x}_i^a \phi\left(x_j^a\right)\right]\right\rangle_{\mathbf{J}^0}\left\langle\exp \left[i \frac{\sqrt{\beta}}{N} \delta \sum_{i j} u_i \sum_a \hat{x}_i^a \phi^2\left(x_j^a\right)\right]\right\rangle_{\mathbf{u}} \\
    = & \exp \left[-\frac{1}{2} \frac{\beta g^2}{N} \sum_{i j}\left(\sum_a \hat{x}_i^a \phi\left(x_j^a\right)\right)^2-\frac{1}{2} \frac{\beta}{N^2} \delta^2 \sum_i\left(\sum_j \sum_a \hat{x}_i^a \phi^2\left(x_j^a\right)\right)^2\right] \\
    = & \exp \left[-\frac{1}{2} \frac{\beta g^2}{N} \sum_{i j} \sum_{a b} \hat{x}_i^a \hat{x}_i^b \phi\left(x_j^a\right) \phi\left(x_j^b\right)-\frac{1}{2} \frac{\beta}{N^2} \delta^2 \sum_{i j k} \sum_{a b} \hat{x}_i^a \hat{x}_i^b \phi^2\left(x_j^a\right) \phi^2\left(x_k^b\right)\right] \\
    = & \exp \left[-\frac{1}{2} \beta g^2 \sum_i \sum_{a b} \hat{x}_i^a \hat{x}_i^b Q^{a b}-\frac{1}{2} \beta \delta^2 \sum_i \sum_{a b} \hat{x}_i^a \hat{x}_i^b Q^{a a} Q^{b b}\right],
    \end{aligned}
\end{equation}
where the replica overlap matrix $ Q^{a b}=\frac{1}{N} \sum_i \phi\left(x_i^a\right) \phi\left(x_i^b\right),
$ has to be introduced through an integral of Dirac delta functions:
\begin{equation}
    \begin{aligned}
    1 & =\prod_{a \leq b} \int d Q^{a b} \delta\left(Q^{a b}-\frac{1}{N} \sum_i \phi\left(x_i^a\right) \phi\left(x_i^b\right)\right)  \\
    & =\int \frac{d \mathbf{Q} d \hat{\mathbf{Q}} }{2 \pi} \exp \left[-i \sum_{a \leq b} Q^{a b} \hat{Q}^{a b}+i \frac{1}{N} \sum_i \sum_{a \leq b} \hat{Q}^{a b} \phi\left(x_i^a\right) \phi\left(x_i^b\right)\right] \\
    & =\int \frac{d \mathbf{Q} d \hat{\mathbf{Q}} }{2 \pi i / N} \exp \left[-N \sum_{a \leq b} Q^{a b} \hat{Q}^{a b}+\sum_i \sum_{a \leq b} \hat{Q}^{a b} \phi\left(x_i^a\right) \phi\left(x_i^b\right)\right].
    \end{aligned}
\end{equation}

The $n$-th moment of the partition function can then be written as
\begin{equation}
    \begin{aligned}
    \left\langle Z_\ell^n\right\rangle \propto & \int d \mathbf{x} D \hat{\mathbf{x}} d \mathbf{Q} d \hat{\mathbf{Q}}  \exp \left[-i \sqrt{\beta} \sum_{i a} x_i^a \hat{x}_i^a-\beta \eta \sum_{i a}\left(x_i^a\right)^2-\frac{1}{2} g^2 \beta \sum_{a b} Q^{a b}\left(\sum_i \hat{x}_i^a \hat{x}_i^b\right)\right.\\
    & \left.-\frac{1}{2} \beta \delta^2  \sum_{a b}  Q^{a a} Q^{b b}\left(\sum_i \hat{x}_i^a \hat{x}_i^b\right)
    -N \sum_{a \leq b} Q^{a b} \hat{Q}^{a b}+\sum_i \sum_{a \leq b} \hat{Q}^{a b} \phi\left(x_i^a\right) \phi\left(x_i^b\right)\right] \\
    & \propto \int d \mathbf{Q} d \hat{\mathbf{Q}}  \exp \left[N\left(-\sum_{a \leq b} Q^{a b} \hat{Q}^{a b}+G\right)\right],
    \end{aligned}
\end{equation}
where the model-dependent action reads
\begin{equation}
    \begin{aligned}
    G= & \ln \int d \mathbf{x} D \hat{\mathbf{x}} \exp \left[-i \sqrt{\beta} \sum_a x^a \hat{x}^a-\beta \eta \sum_a\left(x^a\right)^2-\frac{1}{2} g^2 \beta \sum_{a b} Q^{a b} \hat{x}^a \hat{x}^b+\sum_{a \leq b} \hat{Q}^{a b} \phi\left(x^a\right) \phi\left(x^b\right)\right. \\
    & \left.-\frac{1}{2} \beta \delta^2  \sum_{a b}  Q^{a a} Q^{b b}\left( \hat{x}^a \hat{x}^b\right)\right].
    \end{aligned}
\end{equation}

We then adopt the RS ans\"atz once again. $ Q^{a b}=q \delta_{a b}+Q\left(1-\delta_{a b}\right)$. The averaged replicated partition function yields a neat form:
\begin{equation}
    \left\langle Z_\ell^n\right\rangle \propto \int(d Q d \hat{Q} d q d \hat{q}) \exp \left[-N\left(\frac{n(n-1)}{2} Q \hat{Q}+n q \hat{q}\right)\right] \exp [N G],
\end{equation}
where $G$ is defined below: 
\begin{equation}
    \begin{aligned}
    G= & \ln \int d \mathbf{x} D \hat{\mathbf{x}} \exp \left[-i \sqrt{\beta} \sum_a x^a \hat{x}^a-\beta \eta \sum_a\left(x^a\right)^2-\frac{1}{2} g^2 \beta\left(Q\left(\sum_a \hat{x}^a\right)^2+(q-Q) \sum_a\left(\hat{x}^a\right)^2\right)\right. \\
    & \left.+\frac{1}{2}\left(\hat{Q}\left(\sum_a \phi\left(x^a\right)\right)^2+(2 \hat{q}-\hat{Q}) \sum_a \phi^2\left(x^a\right)\right)-\frac{1}{2} \beta \delta^2 q^2 \left( \sum_{a} \hat{x}^a \right)^2 \right].
    \end{aligned}
    \end{equation}
   Notice that $G$ can be further simplified by linearizing the quadratic terms, i.e., by reversely applying a Gaussian integral identity $\int Dt e^{bt}=e^{b^2/2}$. We then arrive at the following result:
   \begin{equation}
    \begin{aligned}
    G = & \ln \int d \mathbf{x} D \hat{\mathbf{x}}(D u D v) \exp \left[-i \sqrt{\beta} \sum_a x^a \hat{x}^a-\beta \eta \sum_a\left(x^a\right)^2-\frac{1}{2} g^2 \beta(q-Q) \sum_a\left(\hat{x}^a\right)^2\right. \\
    & \left.+\frac{1}{2}(2 \hat{q}-\hat{Q}) \sum_a \phi^2\left(x^a\right)+i \sqrt{\beta g^2 Q + \beta \delta^2 q^2} v \sum_a \hat{x}^a+\sqrt{\hat{Q}} u \sum_a \phi\left(x^a\right)\right] \\
    = & \ln \int(D u D v)\left(I_1\right)^n,
    \end{aligned}
\end{equation}
where the shorthand for the one-dimensional integral $I_1$ reads,
\begin{equation}
    \begin{aligned}
    &I_1  =\int(d x D \hat{x}) \exp \left[-i \sqrt{\beta} x \hat{x}-\beta \eta x^2-\frac{1}{2} g^2 \beta(q-Q) \hat{x}^2+\frac{1}{2}(2 \hat{q}-\hat{Q}) \phi^2(x) \right. \\
    & \left.+i \sqrt{\beta g^2 Q + \beta \delta^2 q^2} v \hat{x}+\sqrt{\hat{Q}} u \phi(x)\right] \\
    & =\int d x \exp \left[-\beta \eta x^2+\frac{1}{2}(2 \hat{q}-\hat{Q}) \phi^2(x)+\sqrt{\hat{Q}} u \phi(x)\right] \int D \hat{x} \exp \left[-i \sqrt{\beta} x \hat{x} \right. \\
    &\left.-\frac{1}{2} g^2 \beta(q-Q) \hat{x}^2+i \sqrt{\beta g^2 Q + \beta \delta^2 q^2} v \hat{x}\right] \\
    & =\int d x \frac{1}{\sigma} \exp \left[-\beta \eta x^2+\frac{1}{2}(2 \hat{q}-\hat{Q}) \phi^2(x)+\sqrt{\hat{Q}} u \phi(x)- \frac{1}{2\sigma^2}\left(\sqrt{\beta g^2 Q + \beta \delta^2 q^2} v-\sqrt{\beta} x\right)^2\right],
    \end{aligned}
\end{equation}
where $\sigma \equiv \sqrt{1+g^2 \beta(q-Q)}$.

The $G$ function can be finally written in a neat form as follows:
\begin{subequations}
\begin{align}
    G&=-n \ln \sigma+\ln \int D u D v I^n,\\
    I &\equiv \int d x e^{\mathcal{H}(x)},
    \end{align}
\end{subequations}
where the single-variable effective Hamiltonian for the feedback learning is given below.
\begin{equation}
    \begin{aligned}
    \mathcal{H}(x) & \equiv -\beta \eta x^2+\frac{1}{2}(2 \hat{q}-\hat{Q}) \phi^2(x)+\sqrt{\hat{Q}} u \phi(x)- \frac{1}{2\sigma^2}\left(\sqrt{\beta g^2 Q + \beta \delta^2 q^2} v-\sqrt{\beta} x\right)^2.
    \end{aligned}
\end{equation}
Hence, applying the replica trick, we manage to complete the quenched disorder average and get the free energy.
\begin{equation}
    -\beta f=\frac{1}{2} Q \hat{Q}-q \hat{q}-\ln \sigma+\int D u D v \ln I.
\end{equation}

We are interested in the zero-temperature limit for the same reason claimed before. To get a physically meaningful free energy in this limit, we have
to adopt the following scaling behavior.
\begin{equation}
    \begin{aligned}
    & (q-Q) \rightarrow \frac{\chi}{\beta} ,\\
    & (2 \hat{q}-\hat{Q}) \rightarrow \beta \hat{\chi}, \\
    & q \rightarrow q ,\\
    & \hat{q} \rightarrow \beta^2 \hat{q} .\\
    \end{aligned}
\end{equation}
One can thus get the scaling behavior of the effective Hamiltonian as follows,
\begin{equation}
    \begin{aligned}
    \mathcal{H}(x) & \equiv-\beta \eta x^2+\frac{1}{2}(2 \hat{q}-\hat{Q}) \phi^2(x)+\sqrt{\hat{Q}} u \phi(x)- \frac{1}{2\sigma^2}(\sqrt{\beta g^2 Q + \beta \delta^2 q^2} v-\sqrt{\beta} x)^2 ,\\
    & \Rightarrow -\beta \eta x^2+\frac{1}{2}\beta \hat{\chi} \phi^2(x)+ \beta \sqrt{2 \hat{q} } u \phi(x)- \frac{\beta}{2\sigma^2}(\sqrt{ g^2 q +\delta^2 q^2} v- x)^2 \equiv \beta \mathcal{H}_0(x),
    \end{aligned}
\end{equation}
Here $\sigma=\sqrt{1+\beta g^2(q-Q)} \rightarrow \sqrt{1+g^2 \chi}$.

The free energy can be derived by following similar steps.
\begin{equation}
    \begin{aligned}
    -\beta f & =\frac{1}{2} Q \hat{Q}-q \hat{q}-\ln \sigma+\int D u D v \ln \int d x e^{\mathcal{H}(x)} \\
    & =-\frac{1}{2}\left[q(2 \hat{q}-\hat{Q})+\hat{Q}(q-Q)\right] -\ln \sigma+\int D u D v \ln \int d x e^{\mathcal{H}(x)} \\
    & \Rightarrow -\frac{1}{2} \beta(q \hat{\chi}+2 \hat{q} \chi) -\ln \sigma +\int D u D v \ln \int d x e^{\beta \mathcal{H}_0(x)} ,\\
\end{aligned}
\end{equation}
which gives rise to the zero-temperature free energy as follows.
 \begin{equation}
  -f  =-\frac{1}{2}(q \hat{\chi}+2 \hat{q} \chi) + \int(D u D v) \mathcal{H}_0\left(x^*\right),
  \end{equation}
where $x^*=\operatorname{argmax}_x \mathcal{H}_0(x)$, and
\begin{equation}
    \mathcal{H}_0(x)=- \eta x^2+\frac{1}{2} \hat{\chi} \phi^2(x)+  \sqrt{2 \hat{q} } u \phi(x)- \frac{1}{2\sigma^2}(\sqrt{ g^2 q +\delta^2 q^2} v- x)^2,
\end{equation}
From the vanishing gradients of the free energy with respect to associated order parameters, we get the SDEs below:
\begin{equation}
    \begin{aligned}
    q & =\left[\phi^2\left(x^*\right)\right] ,\\
    \chi & =\frac{1}{\sqrt{2 \hat{q}}}\left[u \phi\left(x^*\right)\right] ,\\
    \hat{q} & = \frac{g^2}{2 \sigma^4}\left(g^2 q + \delta^2 q^2+\left[\left(x^*\right)^2\right]-2 \sqrt{g^2 q + \delta^2 q^2}\left[v x^*\right]\right) ,\\
    \hat{\chi} & = -\frac{g^2}{\sigma^2}-\frac{2 q \delta^2}{\sigma^2}+ \frac{g^2 + 2q \delta^2}{\sigma^2\sqrt{g^2 q+\delta^2 q^2}}\left[v x^*\right].
    \end{aligned}
\end{equation}

Because in the feedback learning, the response function can not be retrieved from the vanilla replica calculation, we can use the moment generating function to
derive the response function. We thus modify the quasi-potential as
\begin{equation}
    E_\ell(\bx)=\frac{1}{2} \sum_i\left(-x_i+\sum_j J_{i j} \phi\left(x_j\right)+h_i\right)^2+\eta \sum_i x_i^2 + \gamma \sum_i \phi(x_i),
\end{equation}
where $\gamma$ is a Lagrange parameter for the population current, and $\mathbf{h}$ is a weak external input vector used to trigger the response of the population activity. Introducing $n$ replicated dynamical states $\{\bx^a\}_{a=1}^{n}$, one can calculate the quenched average of the replicated partition function as follows,
\begin{equation}
    \begin{aligned}
    & \left\langle Z_\ell^n  \right\rangle  = \left\langle \int d \mathbf{x} \exp \left[-\beta\left(\frac{1}{2} \sum_{i a}\left(-x_i^a+\sum_j J_{i j} \phi\left(x_j^a\right)+h^a_i\right)^2+\eta \sum_a\left\|\mathbf{x}^a\right\|^2+ \gamma \sum_{i a} \phi(x^a_i)\right)\right] \right\rangle\\
    & = \left\langle \int d \mathbf{x} D \hat{\mathbf{x}} \exp \left[i \sqrt{\beta} \sum_{i a} \hat{x}_i^a\left(-x_i^a+\sum_j J_{i j} \phi\left(x_j^a\right)+h^a_i\right)-\beta \eta \sum_a\left\|\mathbf{x}^a\right\|^2- \beta \gamma \sum_{i a} \phi(x^a_i)\right] \right\rangle,
    \end{aligned}
\end{equation}
where $\langle\cdot\rangle$ indicates the disorder average over the untrained coupling and feedback weights.

Following a similar procedure as before, one obtains the free energy in the zero temperature limit.
\begin{equation}
    -f  =-\frac{1}{2}(q \hat{\chi}+2 \hat{q} \chi) + \int D u D v \mathcal{H}_0\left(x^*\right),
\end{equation}
where $x^*=\operatorname{argmax}_x \mathcal{H}_0(x)$, and
\begin{equation}
    \mathcal{H}_0(x)=- \eta x^2 - \gamma \phi(x) +\frac{1}{2} \hat{\chi} \phi^2(x)+  \sqrt{2 \hat{q} } u \phi(x)- \frac{1}{2\sigma^2}(\sqrt{ g^2 q +\delta^2 q^2} v + h- x)^2.
\end{equation}
Hence, the mean population activity $\langle\phi\rangle\equiv\frac{1}{N}\sum_i\phi_i$ can be obtained from the generating function.
\begin{equation}
   \left\langle\phi\right\rangle =\left.\frac{\partial\left( - f \right)}{\partial \left(-\gamma\right)}\right|_{\gamma = 0}.
\end{equation}
Then the response function can be calculated by definition.
\begin{equation}\label{eqr}
    \begin{aligned}
    &r =\left.\frac{\partial\left\langle\phi\right\rangle}{\partial h}\right|_{h \rightarrow 0}=\left.\frac{\partial\left( - f \right)}{\partial h \partial \left(-\gamma\right)}\right|_{h = 0,\gamma = 0} \\
    & = -\frac{\sqrt{ g^2 q +\delta^2 q^2}}{\sigma^2}\left[v \phi\left(x^*\right)\right]+\frac{1}{\sigma^2}\left[x^* \phi\left(x^*\right)\right]-\frac{1}{\sigma^2}[x^*][\phi(x^*)].
    \end{aligned}
\end{equation}
To derive Eq.~\eqref{eqr}, we assume a finite temperature and finally send the temperature to zero. Because of symmetry in the effective Hamiltonian, the last term in the last equality of Eq.~\eqref{eqr} vanishes.
This response function characterizes how responsive one network state is to a weak external perturbation. Note that in our previous work~\cite{Qiu-2024}, 
apart from the overlap matrix, there appears another response matrix. The diagonal element is exactly the quantity $r$ here. However, the off-diagonal element explains how two dynamical states impact each other, which may be related to psychedelics~\cite{Geo-2024}, but we do not attempt to discuss this phenomenon in this paper.

\subsection{Homeostatic learning}
For the homeostatic learning, the procedure of replica calculation is quite similar to that in the Hebbian learning. One distinct part arises in the quenched disorder-average detailed below.
\begin{equation}
    \begin{aligned}
    & \left\langle\exp \left[i \sqrt{\beta} \sum_{i j} J_{i j} \sum_a \hat{x}_i^a \phi\left(x_j^a\right)\right]\right\rangle \\
    = & \left\langle\exp \left[i \sqrt{\beta} \sum_{i j} J_{i j}^0 \sum_a \hat{x}_i^a \phi\left(x_j^a\right)\right]\right\rangle \exp \left[-i \frac{k \sqrt{\beta}}{N} \sum_{i j} \sum_a \hat{x}_i^a \phi\left(x_i^a\right) \phi^2\left(x_j^a\right) \right. \\
    & \left. +\frac{i \sqrt{\beta} k}{N} r_{\rm tg} \sum_{i j} \sum_{a} \hat{x}_i^a  \phi^2\left(x_j^a\right)\right] \\
    = & \exp \left[-\frac{1}{2} \frac{\beta g^2}{N} \sum_{i j}\left(\sum_a \hat{x}_i^a \phi\left(x_j^a\right)\right)^2-i \frac{k \sqrt{\beta}}{N} \sum_{i j} \sum_a \hat{x}_i^a \phi\left(x_i^a\right) \phi^2\left(x_j^a\right) \right. \\
    & \left.+\frac{i \sqrt{\beta} k}{N} r_{\rm tg} \sum_{i j} \sum_{a} \hat{x}_i^a  \phi^2\left(x_j^a\right)\right] \\
    = & \exp \left[-\frac{1}{2} \frac{\beta g^2}{N} \sum_{i j} \sum_{a b} \hat{x}_i^a \hat{x}_i^b \phi\left(x_j^a\right) \phi\left(x_j^b\right)-i \frac{k \sqrt{\beta}}{N} \sum_{i j} \sum_a \hat{x}_i^a \phi\left(x_i^a\right) \phi^2\left(x_j^a\right) \right. \\
    & \left. +\frac{i \sqrt{\beta} k}{N} r_{\rm tg} \sum_{i j} \sum_{a} \hat{x}_i^a  \phi^2\left(x_j^a\right)\right] \\
    = & \exp \left[-\frac{1}{2} \beta g^2 \sum_i \sum_{a b} \hat{x}_i^a \hat{x}_i^b Q^{a b}-i kN \sqrt{\beta} \sum_a R^a Q^{a a} +i \sqrt{\beta} k r_{\rm tg} \sum_{i} \sum_{a} \hat{x}_i^a  Q^{a a}\right].
    \end{aligned}
\end{equation}
The remaining steps are similar to those in Sec.~\ref{app-a1}. The final expressions of free energy in both finite and zero temperatures are summarized in the main text, together with the associated saddle-point equations in the zero temperature limit.

\section{Numerical details of solving SDEs}\label{app-b}
To iteratively solve the saddle point equations, we first initialize the order parameters, and then generate $M=100\,000$ Monte Carlo samples and identify the global maximum of $\mathcal{H}_0(x)$, i.e., $x^*$ for each pair of $(u, v)$ samples. Then we update the order parameters according to the following procedure until the convergence (i.e., $\left|\mathcal{O}_{t+1}-\mathcal{O}_t\right|<10^{-3}$) is achieved. To speed up convergence, we use the following damping step:
\begin{equation}
  \mathcal{O}_{t+1}=\alpha \mathcal{O}_t+(1-\alpha) f\left(\mathcal{O}_t\right),
\end{equation} 
Here, $\alpha=0.2$ is the damping parameter. The pseudocode is given in Algorithm~\ref{alg}.
In practice, we also use the convergent order parameters at a smaller value of $g$ to initialize the iteration of the SDEs for a larger value of $g$.
This trick is useful to mitigate numerical instability. 

\begin{algorithm}[H]
    \caption{SDE solver}\label{alg}
    \begin{algorithmic}[1]
    \Require $g$, initial values of $\mathcal{O}\in\{q,\chi,r\}$ and a damping factor $\alpha$
    \Ensure convergent values of $\mathcal{O}$
    \Repeat 
        \State generate Gaussian samples $u,v$
        \State find $x^{*}$ by the golden section search for the function $\mathcal{H}_0(x)$
        \State calculate the average $[\cdot]$
        \State $\mathcal{O}_{t+1}\leftarrow\alpha\mathcal{O}_{t}+\left(1-\alpha\right)f\left(\mathcal{O}_{t}\right)$, where $f\left(\mathcal{O}\right)$ is the right hand side of the saddle-point equations (see the main text)
    \Until convergence
    \end{algorithmic}
\end{algorithm}

\section{The FORCE Learning Algorithm}\label{app-c}
\blue{The FORCE (First-Order Reduced and Controlled Error) learning algorithm is a supervised online method for training recurrent neural networks while stabilizing their potentially chaotic dynamics. It is based on the recursive least squares (RLS) framework and updates weights rapidly to ensure that the network output tracks a target signal~\cite{Abbott-2009,Huang-2022}.

The network output is given by
\begin{equation}
z(t) = \mathbf{w}^\top \mathbf{r}(t),
\end{equation}
where $\mathbf{r}(t) = \tanh(\mathbf{x}(t))$ are the firing rates of the recurrent units. The error signal is defined as
\begin{equation}
e(t) = z(t) - f(t),
\end{equation}
with $f(t)$ the desired target output.

The output weights are updated at each learning step $\Delta t$ according to
\begin{equation}
\mathbf{w}(t) = \mathbf{w}(t-\Delta t) - e(t)\,\mathbf{P}(t)\,\mathbf{r}(t),
\end{equation}
where $\mathbf{P}(t)$ is an adaptive estimate of the inverse correlation matrix of $\mathbf{r}(t)$. Instead of recomputing the inverse explicitly, $\mathbf{P}(t)$ is updated recursively using the Woodbury matrix identity:
\begin{equation}
\mathbf{P}(t) =
\mathbf{P}(t-\Delta t) -
\frac{
\mathbf{P}(t-\Delta t)\,\mathbf{r}(t)\mathbf{r}^\top(t)\,\mathbf{P}(t-\Delta t)
}{
1 + \mathbf{r}^\top(t)\,\mathbf{P}(t-\Delta t)\,\mathbf{r}(t)
}.
\end{equation}
The initialization $\mathbf{P}(0)=\alpha^{-1}\mathbf{I}$, and we set $\alpha = 1$ in all experiments.}



\end{document}